\documentclass[fleqn,usenatbib]{mnras}

\usepackage{newtxtext,newtxmath}
\usepackage[T1]{fontenc}
\usepackage{ae,aecompl}

\usepackage{graphicx}	% Including figure files
\usepackage{amsmath}	% Advanced maths commands
\usepackage{amssymb}	% Extra maths symbols
\usepackage{longtable}

\graphicspath{{images/}}
\title[An EAGLE's View of Ex-situ Galaxy Growth]{An EAGLE's View of Ex-situ Galaxy Growth}

\author[T. A. Davison et al.]{
Thomas A. Davison,$^{1,2,3}$\thanks{E-mail: TDavison@uclan.ac.uk}
Mark A. Norris,$^{2}$, Joel L. Pfeffer$^{4}$, Jonathan J. Davies$^{4}$, \newauthor
 Robert A. Crain$^{4}$
\\
$^{1}$Jeremiah Horrocks Institute, University of Central Lancashire, Preston PR1 2HE, UK\\
$^{2}$European Southern Observatory, Karl-Schwarzschild-Strasse 2, D-87548 Garching bei Muenchen, Germany\\
$^{3}$Isaac Newton Group of Telescopes, 38700 Santa Cruz de La Palma, Spain; \\
$^{4}$Astrophysics Research Institute, Liverpool John Moores University, 146 Brownlow Hill, Liverpool L3 5RF, UK \\
}

\date{Accepted XXX. Received YYY; in original form ZZZ}

\pubyear{2020}

\begin{document}
\label{firstpage}
\pagerange{\pageref{firstpage}--\pageref{lastpage}}
\maketitle

\begin{abstract}
Modern observational and analytic techniques now enable the direct measurement of star formation histories and the inference of galaxy assembly histories. However, current theoretical predictions of assembly are not ideally suited for direct comparison with such observational data. We therefore extend the work of prior examinations of the contribution of ex-situ stars to the stellar mass budget of simulated galaxies. Our predictions are specifically tailored for direct testing with a new generation of observational techniques by calculating ex-situ fractions as functions of galaxy mass and morphological type, for a range of surface brightnesses. These enable comparison with results from large FoV IFU spectrographs, and increasingly accurate spectral fitting, providing a look-up method for the estimated accreted fraction. We furthermore provide predictions of ex-situ mass fractions as functions of galaxy mass, galactocentric radius and environment. Using $z=0$ snapshots from the 100cMpc$^3$ and 25cMpc$^3$ EAGLE simulations we corroborate the findings of prior studies, finding that ex-situ fraction increases with stellar mass for central and satellite galaxies in a stellar mass range of 2$\times$10$^{7}$ - 1.9$\times$10$^{12}$ M$_{\odot}$. For those galaxies of mass M$_*$>5$\times$10$^{8}$M$_{\odot}$, we find that the total ex-situ mass fraction is greater for more extended galaxies at fixed mass. When categorising satellite galaxies by their parent group/cluster halo mass we find that the ex-situ fraction decreases with increasing parent halo mass at fixed galaxy mass. This apparently counter-intuitive result may be the result of high passing velocities within large cluster halos inhibiting efficient accretion onto individual galaxies.
%245 words out of 250
\end{abstract}


%%%%%%%%%%%%%%%%%%%%%%%%%%%%%%%%%%%%%%%%%%%%%%%%%%

%%%%%%%%%%%%%%%%% BODY OF PAPER %%%%%%%%%%%%%%%%%%
\section{Introduction}\label{intro}
The recent simultaneous advancement of improved observational technology and stunningly detailed theoretical simulations has significantly expanded and deepened the scientific repertoire available to study galaxy formation. 

One of the key insights inferred from large cosmological simulations of galaxy formation \citep[e.g. the EAGLE, Illustris, Horizon-AGN, and Fire simulations;][]{horizonagn,firesims,Vogelsberger14,vogelsberger2014introducing,crain2015eagle,schaye2014eagle} is that galaxies (particularly massive elliptical galaxies) undergo a `two-phase' process of assembly \cite[see e.g.][]{oser2010two}, in which they initially form a relatively compact core of stars formed in-situ to the main progenitor from infalling cosmological cold gas and gas returned to the ISM by stellar evolution. Following this period (at $z\lesssim3$, \citealt{oser2010two}) simulations indicate continuing accretion of stars from smaller galaxies outside of the virial radius via mergers and tidal stripping, resulting in a build up of ex-situ stars. These simulations indicate that the ex-situ fraction is a strong function of the mass and formation history of the galaxy, with galaxies of M$_*$=10$^9$ M$_{\odot}$ being almost entirely composed of stars formed in-situ, while in contrast $\approx$90\% of the stellar mass of the most massive galaxies (M$_{\odot}$ $>$ 1.7$\times$10$^{11}$ M$_{\odot}$) can have originated in previously distinct progenitors  \citep{oser2010two,lackner2012building,rodriguez2016stellar,pillepich2017first}.

Observational evidence has accumulated to support the two-mode scenario of in-situ and ex-situ stellar assembly from a variety of sources. The most obvious is the observation that many, if not most galaxies, including our own Milky Way, exhibit streams of stars attributable to the remains of tidally disrupted dwarf galaxies \cite[see e.g.][]{belokurov2006field, martinez2008ghost, jennings2015ngc, hood2018origin}. Examination of the distribution and orbits of globular clusters (GCs) has shown that the two GC sub-populations typically found around massive galaxies can be associated with the two phases; metal-rich GCs with the in-situ burst of formation in the main progenitor, and metal-poor GCs with the later accretion of lower mass galaxies hosting lower metallicity GC systems \cite[e.g.][]{Forbes11,Romanowsky12,pota2015sluggs,Beasley18, kruijssen, Fahrion}. Investigations of the radial gradients of stellar populations in massive early-type galaxies shows that the outskirts of such galaxies are composed of older, lower metallicity, alpha-enhanced stars indicative of accretion from lower mass progenitors \cite[e.g.][]{Greene12,LaBarbera12,Martin18}. The same mode of galaxy assembly has also been suggested by independent observational means such as supermassive black hole growth \citep{krajnovic2017two}. Finally, observations indicate that at earlier epochs galaxies were considerably more compact than galaxies of comparable mass today, an observation that can in part be explained by the later addition of accreted stars to the outer regions of galaxies \cite[][]{van20143d,vanDokkum14b}.

Recently, observational technology and analysis techniques have advanced sufficiently that the opportunity now exists to infer in detail the star formation and mass assembly histories of galaxies on an individual basis. This advancement is largely due to three developments; 
\begin{enumerate}
\item The advent of integral field spectrographs with a large field of view ($\gtrsim$ 1 arcminute) \cite[e.g. SAURON at the WHT, GCMS on the 2.7m Harlan J. Smith telescope and MUSE at the VLT,][]{sauron1,hill08,bacon2010muse}.
These and similar instruments enable the simultaneous measurement of high signal-to-noise spectra out to large galactocentric distances, delivering unprecedented spatially-resolved measurements of spectroscopically-derived stellar population parameters across galaxies  \cite[see e.g.][]{guerou2016exploring, mentz2016abundance}. 
\item The development of full spectral fitting methods \cite[e.g. pPXF, STARLIGHT, VESPA etc;][]{cappellari2004parametric,starlight,vespa,Cappellari17} to extract temporally resolved star formation histories from the integrated light measurements made by spectrographs \cite[see e.g.][for demonstrations of the application of this approach]{Onodera12,norris2015extended,ferre-mateu17,Kacharov18,ruiz-lara18}. 
\item The development of techniques to use the information provided by the full star formation and chemical enrichment histories provided by ii) to infer galaxy accretion histories \citep{boecker2019galaxy, kruijssen2019mosaics} or to provide additional constraints on assembly history in combination with kinematic information \citep[e.g. a population-dynamical approach;][]{Poci19}.
\end{enumerate} 

In order to aid the understanding of the huge quantity of information provided by the combination of these technologies and techniques, it is necessary to have more detailed predictions from simulations with which to compare to the observations. To date, most studies of the assembly histories of galaxies have focused on examining the underlying physics from a more theoretical direction \cite[e.g.][]{oser2010two,rodriguez2016stellar,qu2016chronicle}. In this paper we use the EAGLE simulations to develop predictions for the fraction of stars that formed in-situ and ex-situ, in terms of observable quantities most useful to observers e.g. galaxy stellar mass, galaxy type (roughly early or late type), environment (specifically halo mass), and most usefully, surface brightness. This study therefore builds on that of \cite{qu2016chronicle} whose analysis of EAGLE found that typically, galaxies of mass M$_*<$ 10$^{10.5}$M$_{\odot}$ assemble less than 10\% of their mass from ex-situ sources, whilst those of M$_*>$10$^{11}$M$_{\odot}$ typically exhibit ex-situ mass fractions of $\sim$20\% (though with a large scatter in these values).

This paper is organised as follows. In Section 2 we describe the EAGLE simulations we are utilising, as well as the methodology for classifying star particles as in- or ex-situ. In Section 3 we present the results of our investigation of the dependence of in- and ex-situ mass fractions as a function of various observable properties of galaxies, and as a function of galactocentric radius. Furthermore we provide tabulated data for use when comparing to observational studies. In Section 4 we discuss some of the implications of our results in light of expectations and previous studies. Finally in Section 5 we provide some concluding remarks.

\section{Methodology}
\subsection{Simulations Overview}

The EAGLE (Evolution and Assembly of GaLaxies and their Environments) simulations are a suite of cosmological hydrodynamical simulations created with the aim of understanding the coevolution of galaxies and supermassive black holes within a cosmologically representative volume of a standard $\Lambda$ cold dark matter Universe. For a more comprehensive overview of the full suite of simulations see \cite{schaye2014eagle,crain2015eagle}. 

For the present work, we focus on two simulations,  Recal-L025N0752 (hereafter Recal-025) and Ref-L0100N1504 (hereafter Ref-100). The Ref-100 simulation is a periodic volume, 100cMpc on a side, realised with 1504$^3$ dark matter particles and an initially equal number of gas particles, described as `intermediate resolution'. The resultant initial baryonic particle mass is 1.81x10$^6$M$_{\odot}$; the dark matter particle mass is 9.70x10$^6$M$_{\odot}$; the gravitational softening length is 2.66ckpc (co-moving kiloparsecs) limited to a maximum proper length of 0.7pkpc (proper kiloparsecs).

Subgrid feedback parameters were calibrated to ensure $z=0$ reproduction of the galaxy stellar mass relation, disc sizes, and the M$_{BH}$ - M$_*$ relation. Further studies make comparisons to other observational properties, showing close matches \citep{schaye2014eagle,furlong2015evolution,lagos2015molecular}. Detailed descriptions of the calibration procedure and the influence of parameter variation are given by \cite{crain2015eagle}. 

Recal-025 is the higher resolution of the two simulations, and the parameters of the Recal model were re-calibrated to achieve a similarly good match to the observational data as the Reference model. This simulation is 25cMpc on a side, realised with 752$^3$ dark matter and gas particles. The resultant initial baryonic particle mass is 2.26x10$^5$M$_{\odot}$; the dark matter particle mass is 1.21x10$^6$M$_{\odot}$; the gravitational softening length is 1.33ckpc limited to a maximum proper length of 0.35pkpc.

The EAGLE simulations use a Chabrier initial stellar
mass function \cite{chabrier2003galactic} and was run with a modified version of the gadget-3 smoothed particle hydrodynamics code. The subgrid treatments are well explained in \cite{mcalpine2016eagle}.

For our analysis we focus on the final $z=0$ snapshots for both simulations, where the simulated galaxies are comparable to local galaxies \citep[see e.g.][]{schaye2014eagle,Trayford15}. These specific simulations are chosen for their complementary ability to adequately resolve the formation of lower mass (M < 10$^9$ M$_\odot$) galaxies (Recal-025), and to provide sufficient statistics on the properties of rarer high mass galaxies (Ref-100). We limit our analyses to only those galaxies which contain a minimum count of $>$500 bound stellar particles at $z=0$, to ensure reasonable sampling errors on derived properties.

\subsection{In/Ex-Situ Classification}\label{inexclass}

The classification of any simulated star (or dark matter) particle as `in' or `ex'-situ is a matter of some debate in the literature, with several schemes implemented to date \cite[see e.g.][]{oser2010two,pillepich2015,rodriguez2016stellar,qu2016chronicle,Clauwens18,antonela2019}. In this current work we implement a scheme to classify each particle as described below.
 
In common with most modern cosmological simulations, the EAGLE simulations track galaxy merger histories by post-processing halo and subhalo catalogues to produce `merger trees'. The structures are connected at different epochs using the D-Trees algorithm \citep{jiang2014n} which tracks linked particles to identify progenitors and descendants. As galaxies merge, the total mass of each branch forming that galaxy is summed, and the larger branch mass is defined as the `main progenitor'.
 
In our scheme, the subhalo that an initial gas particle is a member of (if any) in the snapshot immediately before conversion to a star particle is identified. Following this, the unique identifier of that progenitor, as determined by the merger tree, is assigned to the particular star particle. If the particle happens to be unbound at that snapshot, the earliest snapshot after star formation in which the particle resides in a subgroup as a member is defined as its host.

We then compare the particle progenitor identifier to the identifier of the subhalo the particle is a member of at $z=0$. If the subhalo at the snapshot prior to star formation is in the main branch of the final galaxy, the particle is considered in-situ. If the progenitor galaxy is not in the main branch it is defined as ex-situ. 

The result is that each of the particles comprising a simulated galaxy at $z=0$ has a flag denoting the nature of its origin. This is performed for all particles and data can be limited to a chosen aperture as usual along with other properties. In addition to this flag, and to the standard properties recorded for each EAGLE particle in its snapshot (e.g. stellar mass, metallicity, coordinates); we also calculate the distance from the galaxy centre (defined by the position of the minimum of the gravitational potential, itself the position of the most bound particle) for each particle . From this information it is trivial to compute spherically averaged stellar half-mass radii r$_{1/2}$. This can be achieved simply by summing stellar particles in order of absolute distance to the galaxy centre, until half the total mass within a chosen aperture (in this study, 100pkpc) has been reached. Using this information it is then possible to search for correlations between in/ex-situ fraction and a host of final galaxy properties. 

Combining the status of the particle origin with other EAGLE particle properties allows us to mask to in-situ or ex-situ stars. We use this to examine ensemble properties as a function of radius, stellar mass and halo mass for individual galaxies. %Figure \ref{fig:amr} shows a single high mass galaxy where we have divided the stellar components by both origin and galactocentric distance to demonstrate the Age-Metallicity relations of these populations.

%\begin{figure*} 
%	\includegraphics[width=0.9\textwidth]{images/eagle_grid.pdf}
%    \caption{The age-metallicity plane for a single massive (M$_{*}$ = 4.5 $\times$ 10$^{11}$ M$_{\odot}$) EAGLE galaxy, separated by the in and ex-situ flag, as well as whether the stars lie inside or outside of the half-mass radius. Metallicity was calculated using the stellar particles only, assuming Z$_{\odot}$=0.012 for consistency with yields from the EAGLE project. The percentage share of the total galactic stellar mass is provided in the lower left corner of each panel. Colour demonstrates point density of particles within a bin. As can be seen for this galaxy ex-situ stars dominate at all radii, but especially outside the half mass radius.}
%    \label{fig:amr}
%\end{figure*} 

\subsection{Galaxy Properties and Classification}\label{galclassorig}

A further property that is of interest to observers is the morphological type of galaxy under study. 
This is especially true in this study as morphological transformation has long been thought to be related to merger history \citep[see e.g.][]{querejeta2015formation,Martin18}, and in fact the addition of ex-situ stars is suggested as the main mode of mass assembly for large early-type galaxies \cite[][]{Martin18}.

Using the co-rotational stellar energy parameter $\kappa_{\rm co}$ made available as part of the EAGLE data release \citep[see e.g.][]{thob2018relationship} for each galaxy in the simulations, we separate galaxies into the `blue sequence' of disky starforming galaxies ($\kappa_{\rm co}$ > 0.4 ) and `red sequence' of more spheroidal passively evolving galaxies ($\kappa_{\rm co}$ < 0.4 ) as advocated by \cite{correa2017relation}. The method is based on earlier development in \cite{sales2010feedback} in which rotational energy is quantified as the fraction of kinetic energy invested in ordered rotation. However the \cite{correa2017relation} method instead uses co-rotational energy, defined as:
\begin{equation}
    \kappa_{co} = \frac{K_{co}^{rot}}{K} = \frac{1}{K}\sum_{i,L_{z,i}>0}\frac{1}{2}m_i\left ( \frac{L_{z,i}}{m_iR_i} \right )^2
	\label{eq:kappa}
\end{equation}
where (L$_{z,i}$ > 0) defines all co-rotating stellar particles within 100 pkpc of the galactic centre, and R$_i$ is the 2-dimensional radius in the plane normal to the rotation axis. This is shown to be efficient in classifying galaxies in the EAGLE simulations.

As an independent check of this morphological classification we also apply a separation based on specific star formation rate (sSFR). By plotting stellar mass vs sSFR for the EAGLE galaxy sample, a boundary sSFR of 0.01 Gyr$^{-1}$ is found to effectively separate the blue sequence and red sequence. Applying this selection only alters the classification of 6.35\% of galaxies, relative to the $\kappa _{co}$ selection, leading to no significant changes of any of our findings. Therefore, to maintain consistency with other studies using EAGLE simulations we use the $\kappa _{co}$ classification in all further analyses.

\subsection{The EAGLE Galaxy Sample}
\begin{figure} 
	\includegraphics[width=\linewidth]{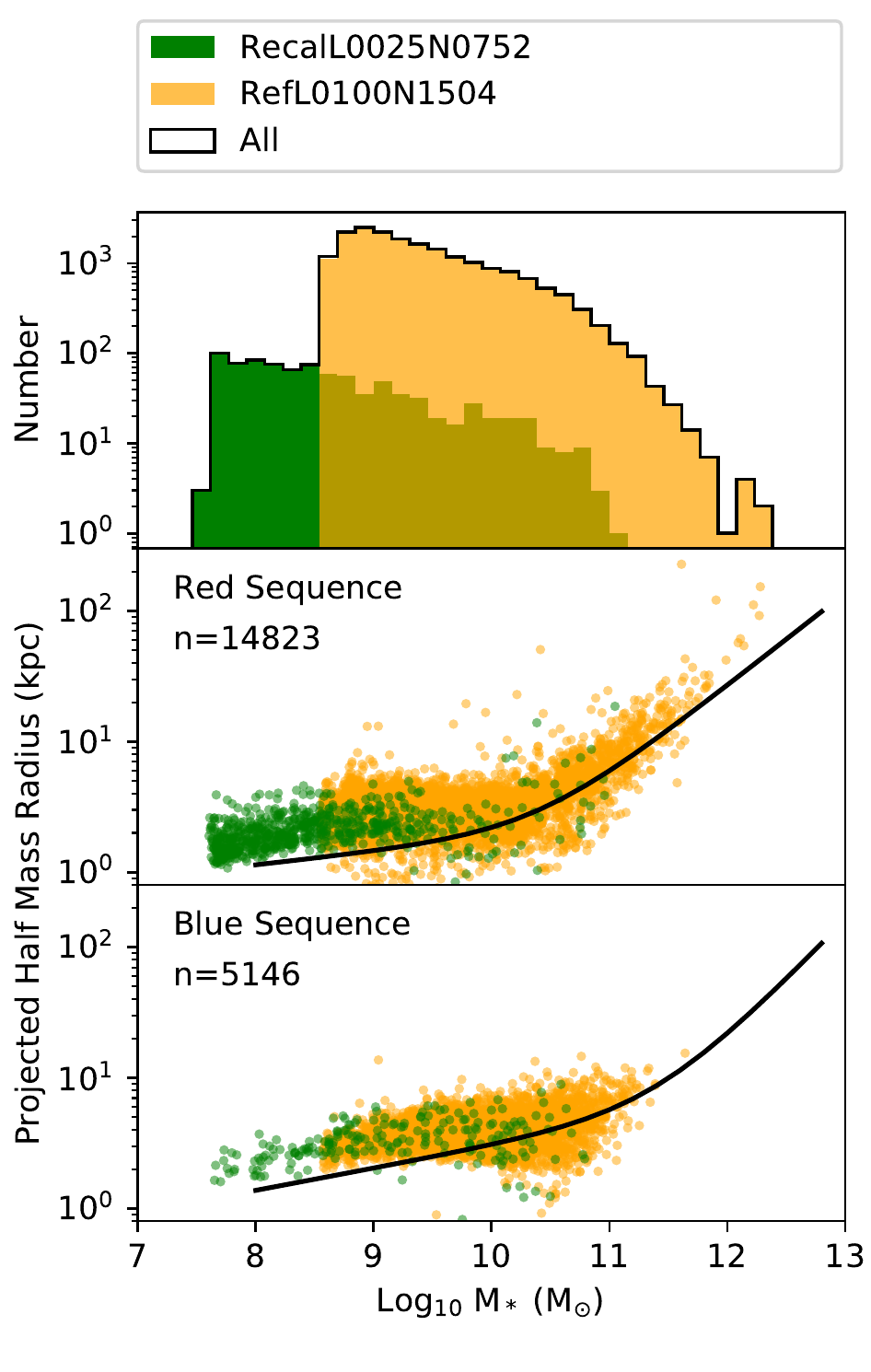}
    \caption{Origin of all objects used in subsequent mass size plots. The upper panel shows a histogram of the mass distribution of simulated galaxies across the (log$_{10}$) stellar mass axis. The centre and lower panels show the same distribution in the mass-size plane, with objects separated into Red sequence and Blue cloud populations, using the $\kappa_{\rm co}$ parameter (see section \ref{galclassorig}). The radius shown is projected half mass radius, averaged over the three orthogonal projections. Fit lines show the double power law mass-size trend of the observed galaxy population in K-band, as described in  \protect\cite{2015MNRAS.447.2603L}. A similar figure for only the Ref-100 simulation can be found in  \protect\cite{furlong2016size} with comparison to measurements from  \protect\cite{van20143d} and  \protect\cite{shen2003size}.}
    \label{hist_trip}
\end{figure}

The histogram in the upper panel of Figure \ref{hist_trip} shows our final sample of EAGLE galaxies. Sharp discontinuities in the apparent distribution of galaxies are the result of the previously described lower particle limit of 500 stellar particles per galaxy. This equates to a minimum stellar mass within 100pkpc of 4.01x10$^7$M$_{\odot}$ and 3.48x10$^8$M$_{\odot}$ at $z=0$ for the Recal-025 and Ref-100 simulations respectively. The most massive galaxies in the two samples have mass 1.12x10$^{11}$M$_{\odot}$ (Recal-025) and 1.93x10$^{12}$M$_{\odot}$ (Ref-100). As can clearly be seen, the Recal-025 and Ref-100 simulation sample different parts of the mass distribution, owing to their differing resolutions and volumes. Recal-025 provides high numbers of low-mass galaxies in the range $\sim$10$^{7.5}$ M$_\odot$-10$^{9.5}$ M$_\odot$, with the Ref-100 simulation taking over (with overlap) to provide galaxies in the range $\sim$10$^{8.5}$ M$_\odot$-10$^{12}$ M$_\odot$. Though Recal-025 provides some small number of galaxies of mass >10$^{9.5}$ M$_\odot$, the fraction is barely noticeable compared to those provided by Ref-100.

The lower two panels of Figure \ref{hist_trip} show the distribution of objects across the mass-size plane. The half-mass radius is determined following the prescription provided in Section \ref{inexclass}. These panels display observational best fit lines from \cite{2015MNRAS.447.2603L} produced for the Galaxy And Mass Assembly (GAMA) survey, in which $M_{*}$ - $r_{1/2}$ relations are built for `red' and `blue' sequence galaxy samples using a double power-law function described by \cite{shen2003size}
\begin{equation}
    r_{1/2} = \gamma \left ( \frac{M_{\star}}{M_{\odot}} \right )^{\alpha}\left (1 + \frac{M_{\star}}{M_0} \right )^{\beta - \alpha}     .
    \label{eqlange}
\end{equation}

We choose the K-band fits due to the fact that the K-band has been shown to be a good stellar mass tracer due to its relative insensitivity to both extinction and the effects of young hot stars which bias mass-to-light measurements in the optical \cite[e.g.][]{McGaugh14,Norris16}. Hence we expect that K should provide the closest match between the simulations which output stellar mass, and the observationally inferred values. The parameters for Equation \ref{eqlange} are provided in  \cite{shen2003size} for K band. For the blue sample, these parameters are $\gamma$ = 0.1, $\alpha$ = 0.16, $\beta$ = 1.00 and M$_0$ = 33.62$\times$10$^{10}$M$_{\odot}$. Parameters for the red sample are $\gamma$ = 0.12, $\alpha$ = 0.1, $\beta$ = 0.78 and M$_0$ = 2.25$\times$10$^{10}$M$_{\odot}$.

As Figure \ref{hist_trip} demonstrates, the simulated galaxies appear to be slightly more extended at fixed mass than real galaxies. This has a number of potential explanations. One is the result of over-efficient feedback within the EAGLE simulations \cite[see e.g.][]{crain2015eagle} which moves galaxies of fixed M$_*$ into haloes of higher M$_{200}$ potentially extending the galaxies. A second explanation is described in \cite{ludlow2019energy}, wherein if a simulation contains particle species of differing mass (such as the lower mass baryonic and higher mass dark-matter particles in EAGLE), 2-body scattering can be artificially inflated. This occurs when higher mass species effectively transfer kinetic energy to lower mass species, artificially raising their energy. This may contribute to the greater extent of the EAGLE galaxies compared to observation that we see in Figure \ref{hist_trip} especially in lower mass galaxies.

\subsection{Summary of Other Quantities}\label{params} Units and conversions are kept consistent with prior EAGLE analysis works. An aperture of 100pkpc is considered for all objects unless otherwise stated. This ensures that galaxies of M$_*>$10$^{11}$M$_\odot$ are properly considered, as demonstrated in figure 4 of \cite{schaye2014eagle}. For subhalo mass units, only the baryonic stellar component is used; however when analysing halo mass (such as in section \ref{reshalmass}) we use the total mass within the corresponding group Crit200 radius, where group Crit200 defines the radius at which mass is equal to 200 times the critical mass of the universe.

In order to use the EAGLE simulations to provide robust predictions for observational studies, we provide estimates of the ex-situ mass fraction as a function of more readily observable quantities. To achieve this, each simulated galaxy is analysed to determine the ex-situ fraction as a function of visual (i.e. V band) surface brightness. The results of this are discussed in Section \ref{obsresults}

To estimate surface brightness quantities, galaxies are projected in the xy plane, effectively randomising the viewing angle relative to the observer. This is intended to give an accurate representation of galaxies compared to real observing conditions. Each galaxy is set at a fiducial distance of 16.5Mpc (the approximate distance to the Virgo cluster). Following this, the luminosity of each particle is calculated by matching each to interpolated stellar mass-to-light relations built from the single stellar population models of \cite{maraston2005evolutionary}.

Derivation of projected ellipticity is completed using the formulae described by \cite{lagos_ellipse}. Here r-band luminosity is used to calculate the projected ellipticity by:

\begin{equation}
    \epsilon = 1 - \sqrt{\frac{b^2}{a^2}}
\end{equation}
Where,
\begin{equation}
    a^2=\frac{\bar{x}^2+\bar{y}^2}{2}+\sqrt{\left (  \frac{{}\bar{x}^2-\bar{y}^2}{2}\right )^2 + \bar{xy}}
\end{equation}    
\begin{equation}
    b^2=\frac{\bar{x}^2+\bar{y}^2}{2}-\sqrt{\left (  \frac{{}\bar{x}^2-\bar{y}^2}{2}\right )^2 + \bar{xy}}
\end{equation}
And,
\begin{equation}
\bar{x}^2 = \frac{\sum_i L_ix_i^2}{\sum_iL_i}
\end{equation}
\begin{equation}
\bar{y}^2 = \frac{\sum_i L_iy_i^2}{\sum_iL_i}
\end{equation}
\begin{equation}
\bar{xy} = \frac{\sum_i L_ix_iy_i}{\sum_iL_i}
\end{equation}

In which L is the r-band luminosity, and x and y are the particle coordinates in the projected frame (with the centre of the galaxy set to 0,0). The position angle of the elliptical fit is given by:
\begin{equation}
    \theta_{PA} = \frac{1}{2}tan^{-1}\left ( \frac{2\bar{xy}}{\bar{x}^2 - \bar{y} ^2} \right)
\end{equation}

Adaptive bins measure the luminosity of each galaxy from the centre outwards. Particles are binned in order of distance (weighted by ellipticity) from the centre, with every 75 particles justifying a new bin. Area and the sum of the luminosity of particles within each bin is calculated, allowing us to build a luminosity profile across the galaxy. 
Ellipse major axis positions, corresponding to the mid-point between integer magnitudes, are read from the luminosity profile (e.g., choosing inner and outer elliptical radii of 22.5 and 23.5 magnitudes for a magnitude of 23). With these boundaries we define new concentric elliptical bins with the total luminosity of particles contained in a bin providing magnitudes per square arcsecond. Using the previously described ex- or in-situ classification of each particle, we can then determine the ex-situ fraction at any desired surface brightness value.

%This method is detailed in Figure \ref{ellip_method}.

%\begin{figure} 
%	\includegraphics[width=\linewidth]{images/Ellip_method.pdf}
%    \caption{\textcolor{blue}{Surface brightness determination methodology of a randomly selected Recal-025 galaxy of stellar mass 1.01$\times 10^{11}$M$_{\odot}$. The upper panel shows elliptical bins used to calculate the surface brightness profile. Boundaries expand from the centre, binning particles from the inside out with respect to the galaxy ellipticity. Coloured points show the distance of a given stellar particle from the closest point on the ellipse boundary for an ellipse with major axis 100kpc. The lower panel shows the surface brightness profile calculated from the binning, in units of apparent magnitudes per square arcsecond. Blue points show the average brightness within each bin, approximated by a function overlaid as a solid line. Integer magnitudes, and the major axis diameter at which these magnitudes occur, are marked with dashed lines. This method is used for all galaxies in the sample.}}
%    \label{ellip_method}
%\end{figure}

\section{Results}\label{results}
\subsection{Ex-situ Fraction as a Function of Galaxy Properties}
\subsubsection{Ex-situ Fraction as a Function of Galaxy Mass}\label{ex_sit gal mass}

In order to examine the reliability of our methodology, we first examine the dependence of ex-situ fraction on various other properties. This allows us to compare to the results of previous studies, and therefore provides confidence in the reliability of our later predictions.

\begin{figure*}
    \includegraphics[width=\linewidth]{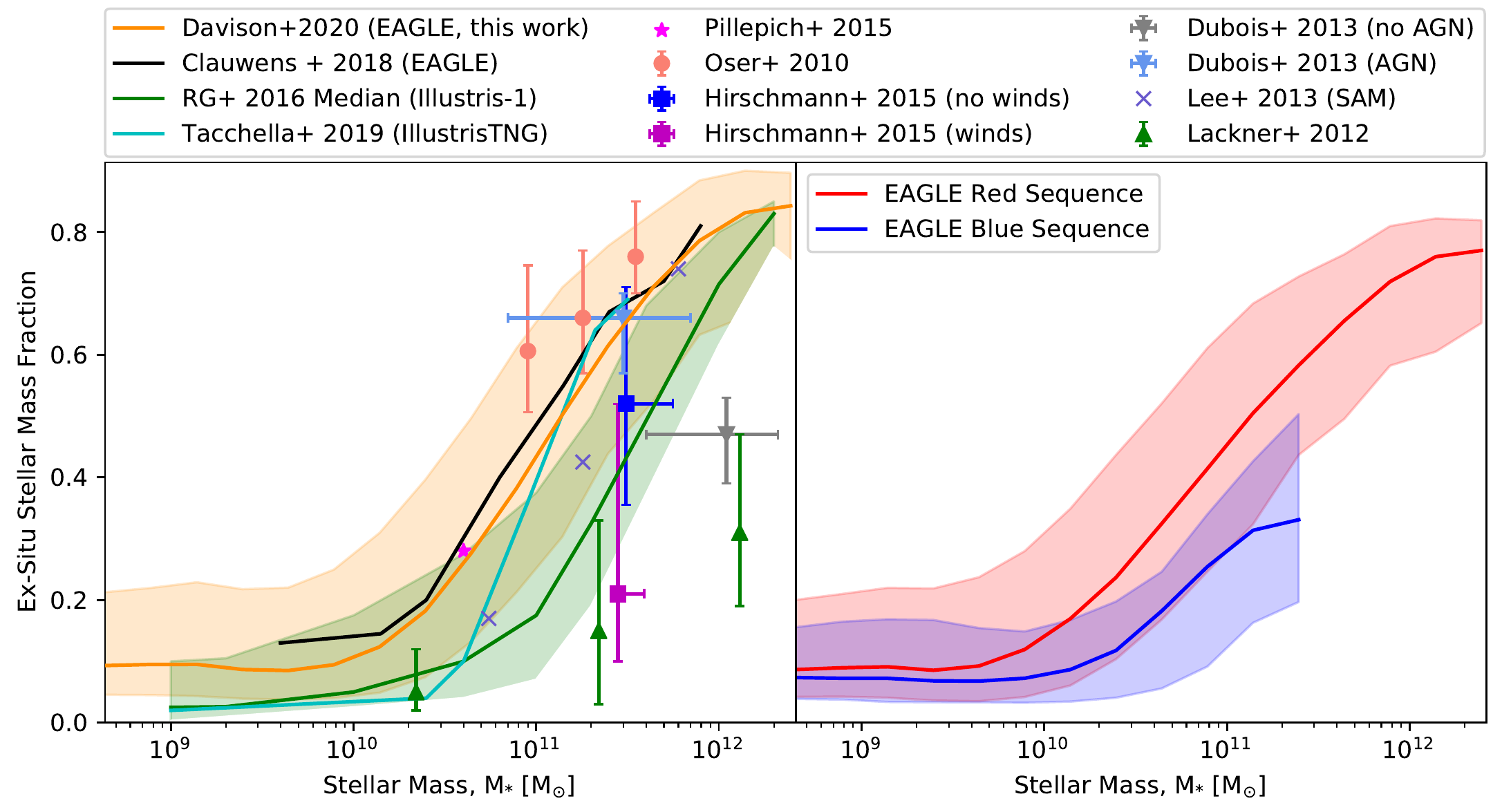}
    \caption{This figure (modelled after figure 3 from  \protect\cite{rodriguez2016stellar}) shows ex-situ fraction as a function of stellar mass and adds z=0 EAGLE data for comparison. EAGLE data consists of a sample built from the Recal-025 and Ref-100 simulations as detailed in Figure \ref{hist_trip}. The left panel compares the full sample of EAGLE galaxies to previous studies. The right panel separates the EAGLE sample into blue sequence and red-sequence components using the $\kappa_{\rm co}$ parameterisation. Shaded regions show 1$\sigma$ percentile scatter in the ex-situ plane. To reproduce accurately the \protect\cite{rodriguez2016stellar} figure, median ex-situ fraction is calculated with 0.25 dex mass bins with a minimum bin population of 8 (the highest mass bin) and a maximum bin population of 3172. Only particles inside of 100pkpc are considered.}
    \label{repo}
\end{figure*}

Figure \ref{repo} displays the stellar mass dependence of the ex-situ fraction for our EAGLE sample along with a selection of other literature sources. The layout follows that of Figure 3 of \cite{rodriguez2016stellar} in which the same analysis is performed on simulated galaxies from the Illustris simulations \citep{genel2014introducing,vogelsberger2014introducing, nelson2015d}. 

As can be seen in Figure \ref{repo} our sample exhibits the same overall trend as found by previous studies \citep{oser2010two,rodriguez2016stellar,qu2016chronicle,Clauwens18}, in that ex-situ fraction increases with increasing galaxy mass. In common with \cite{qu2016chronicle}, we find that EAGLE galaxies display systematically higher ex-situ fractions than Illustris \citep{rodriguez2016stellar} at all stellar masses, though the two samples are consistent within their mutual scatters. IllustrisTNG displays almost identical results as EAGLE at high mass (>2$\times$10$^{11}$ M$_\odot$) \citep{tacchella}. Our sample also displays only marginally lower ex-situ fractions at fixed mass than either \cite{oser2010two} or \cite{Clauwens18}. The analysis of \cite{Clauwens18} makes use of the same Ref-100 EAGLE simulation as this work, although it applies an independent method of classifying in- and ex-situ fraction. The fact that two independent methods for determining ex-situ fraction agree so well, when looking at the same simulation, is encouraging.

The origins of the differences between the ex-situ mass fractions of EAGLE and Illustris are explored by \cite{qu2016chronicle}, with the preponderance of the differences likely caused by differing stellar and AGN feedback treatment. Minor differences between values from our analysis and those of \cite{Clauwens18} can be attributed to the systematic uncertainty associated with using differing methods to classify the origin of stars within a galaxy, such as assigning the particle to a progenitor galaxy prior to or after star formation occurs, as well as a difference in samples examined, with Clauwens et al. not including Recal-025 as we do. The fact that our results agree so closely with those found by several independent analyses of several independent simulations gives confidence that our sample is reliably capturing the underlying behaviour.

The right hand panel of Figure \ref{repo} shows the same data but now split by our morphology classification. Unfortunately the number of `blue sequence' galaxies drops off dramatically above 10$^{11}$ M$_\odot$, which is where the significant increase in ex-situ fraction with mass becomes apparent for the full sample. However, within the stellar mass range where both galaxy types have sufficient numbers for analysis (M$_*$ $<$ 10$^{11}$M$_{\odot}$), there is a statistically significant difference in the two distributions (the Kolomogorov-Smirnov test gives a p-value of p<10$^{-6}$, therefore the hypothesis that the samples come from the same distributions can be rejected). This difference is characterised in the sense that blue sequence galaxies have a lower ex-situ fraction compared to red sequence galaxies of the same mass. This finding is consistent with that of \cite[][]{rodriguez2016stellar}, who found a similar trend for Illustris galaxies.

\subsubsection{Ex-Situ Fractions across the Mass-Size Plane}

We next examine the behaviour of the ex-situ fraction across the galaxy mass-size plane. To display this information, data are projected onto mass-size axes with colour denoting the ex-situ fraction. The great variety in the density of the galaxies across the mass-size plane is mitigated using a tessellated hexagonal binning approach. Bins are adjusted such that they remain of equal size on log-log axes. Galaxies falling within each bin are identified, and the median of the ex-situ fraction for all galaxies within the bin is calculated. Bins have a minimum of 3 galaxies, and the most populated bin contains 1007 galaxies.

\begin{figure*} 
	\includegraphics[width=\textwidth]{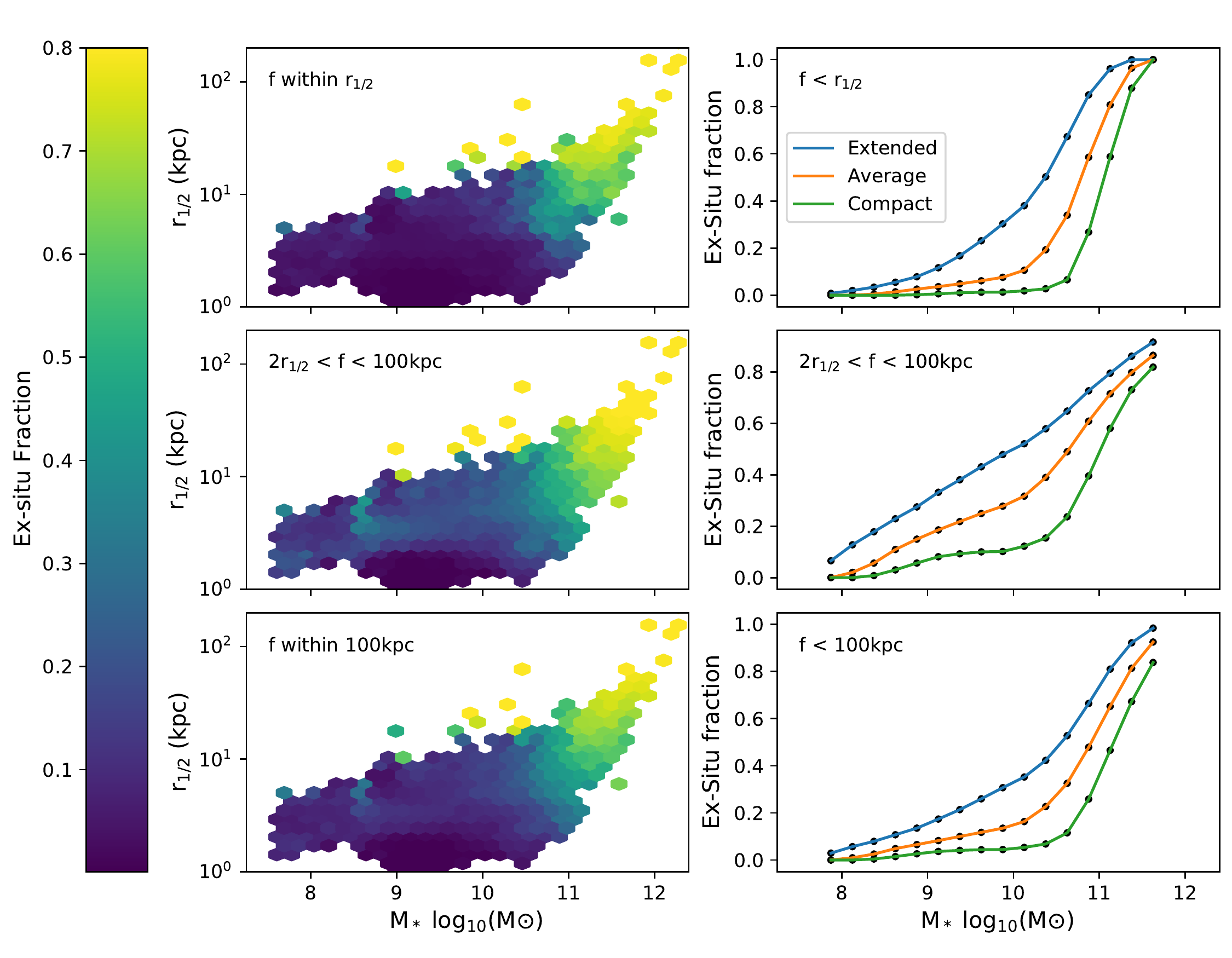}
    \caption{On the left panels, binned z=0 simulated galaxies are shown on the mass-size plane with colouration by ex-situ fraction. The upper-left panel limits to only star particles up to r$_{1/2}$, and the middle-left panel limits to all particles outside of 2r$_{1/2}$(r$<$100pkpc). Finally the lower left panel includes all particles within an aperture of 100pkpc. The panels use a 2D LOESS smoothing function to remove small fluctuations, and to better display the general trend. Colour is given by the median of ex-situ values within a hexagonal bin, with a minimum of 3 objects in a bin, a maximum of 1007 objects in a bin, and an average of 57 objects in a bin. The right-hand panels show mass against ex-situ fraction, when separated into 3 bins of galaxy radius at fixed mass. The `Extended' line collects the upper 33\% of particles spatially on the y axis. The `Average' line represents the central 33\% and the `Compact' line collects the lowest 33\% of objects. The stellar mass axis is binned into 0.25dex bins, with a minimum of 17 objects in a bin.}
    \label{smoothed}
\end{figure*}

We separate particles by radial location within a galaxy, respective to the half mass radius \textit{r$_{1/2}$}. In the upper left panel of Figure \ref{smoothed} we show the ex-situ fraction of particles within \textit{r$_{1/2}$}, and in the centre left panel only particles with a distance from the galaxy centre greater than 2\textit{r$_{1/2}$}(r$<$100pkpc). Finally, in the lower left panel, we show all particles within an aperture of 100pkpc. A LOESS smoothing function \citep{Cappellari2013b} is applied to the image, this implements 2D local adaptive smoothing across the plane \citep{cleveland1988locally} to reduce the influence of outlier points.

Examination of Figure \ref{smoothed} reaffirms the finding from Figure \ref{repo} that as stellar mass increases, so does ex-situ fraction. This trend is clearly present for stars in both the inner and outer regions of galaxies, though clearly ex-situ stars comprise a larger fraction of those found at larger radii. This finding is therefore consistent with those of previous studies which indicate that the majority of ex-situ stars are deposited on the outer regions of galaxies \cite[see e.g.][]{oser2010two,rodriguez2016stellar}.

As a further extension to this analysis we also examine potential relations between ex-situ fraction and galaxy size. To do this at fixed mass we split the galaxy population into three equal (by number of galaxies) bins of galaxy radius, and then calculate the median ex-situ fraction for each bin. These median lines represent the 33\% most extended, 33\% third most compact, and 33\% most intermediate radius galaxies respectively. The radii separated by the bins varies with stellar mass, such that there are always a third of the galaxies at fixed mass within each bin. The right-hand panels of Figure \ref{smoothed} show the result of this procedure. As can be clearly seen there is a distinct separation of the median ex-situ fractions of compact, mid-sized and extended galaxies, for all galaxies at all masses. This separation is apparent both for the stars contained within r$_{1/2}$, and even more so for those beyond 2r$_{1/2}$.

The fact that at fixed mass, more extended galaxies exhibit higher ex-situ mass fraction supports the idea that they are larger because they accreted more stars during generally ``dry" mergers after the initial burst of in-situ star formation that formed the inner regions of the galaxy. The stars accreted during galaxy-galaxy mergers or fly-by interactions will preferentially be deposited at larger radii leading to an increase in the half-mass radius of the galaxy \cite[see e.g.][]{oser2010two,rodriguez2016stellar}. In contrast, gas accreted during ``wet" galaxy interactions can find its way to the inner regions of the galaxy, as it is dissipative. There it can increase the number and density of in-situ stars, potentially even decreasing the half-mass radius of the galaxy \cite[see e.g.][]{Du_2019}. 

\subsubsection{Ex-situ Fraction as a Function of Parent Halo Mass}\label{reshalmass}
\begin{figure} 
	\includegraphics[width=\columnwidth]{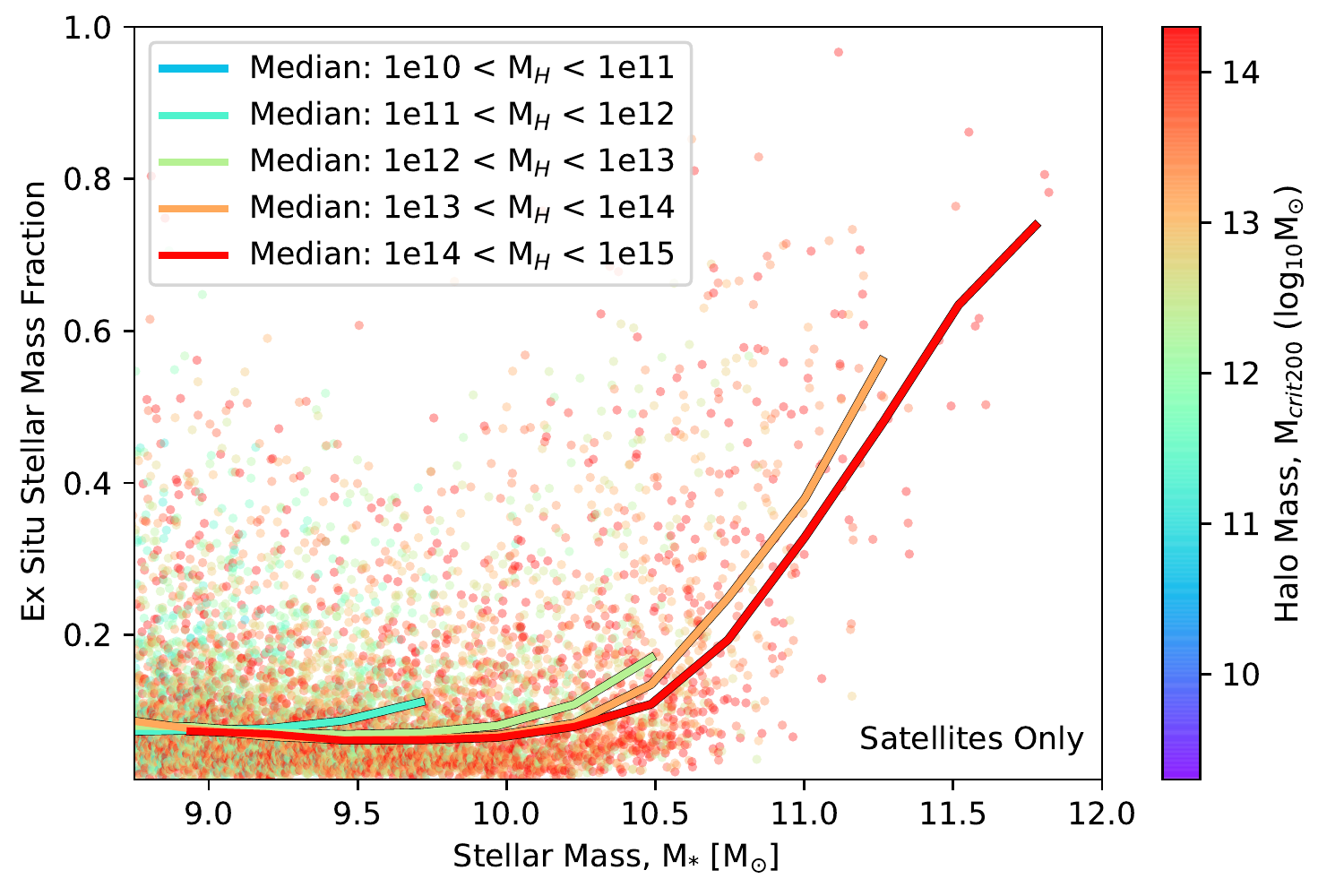}
    \caption{Ex-situ fraction within 100pkpc for all z=0 satellite galaxies with $>$500 stellar particles against stellar mass, and split by total halo mass. Individual objects are coloured by the mass of the halo to which they belong. Solid lines show the median position of the ex-situ fraction for objects within 5 bins of parent halo mass, across 0.25 dex wide bins of stellar mass. There are a minimum of 5 objects per bin, and an average of 220 objects per bin.}
    \label{combo}
\end{figure}

To probe for differences in ex-situ fraction with halo mass, we next limit the sample to include only satellite galaxies (with $>$500 stellar particles) and separate by the mass of the halo they reside in. Figure \ref{combo} shows the result of this analysis for all satellites. 
Analysis was also performed for the centrals only. Though similar effects as discussed below were present, statistical resolution was strained and so the analysis was excluded form this study.
We see in Figure \ref{combo} that ex-situ fraction appears to be highest in objects with a lower parent halo mass, becoming especially distinct at $\sim$M$_*$ $>$ 10$^{10.5}$M$_{\odot}$. At first glance this may seem counter-intuitive, considering that one might reasonably expect more merger activity within denser environments such as those of rich clusters. However, such trends have been hinted at in previous analyses, for example in \citealt[][]{pillepich2017first}, where their Figure 12 shows a similar flattening of the ex-situ fraction increase for the most massive haloes, and indeed a drop in ex-situ fraction for stars in the inner regions of galaxies in the highest mass haloes.

%In order to check the robustness of this seemingly counter-intuitive result we examine the dependence of ex-situ fraction when considering the maximum of the circular velocity curve with z, v$_{\rm max}$(z) (V$_{max}$). We use this rather than halo mass as an independent estimator of galaxy environment. The choice to use the maximum of the circular velocity curve reached at any point in the history of a galaxy ensures that we are not mislead by potential stripping at $z=0$. Such an assumption could potentially lead to temporary increases or decreases in the estimated circular velocity. Therefore, using the maximum of V$_{max}$ (most often occurring at between $z=0$ and $z=1$) by tracking V$_{max}$ back down the merger history and finding the peak value we secure a secondary robust method to characterise the galaxy environment potential. This analysis showed an overall trend consistent with that seen in Figure \ref{combo}. Again we find it is generally the intermediate mass haloes (identified as those with lower peak V$_{max}$) which display maximum ex-situ fraction at fixed galaxy mass. 
Possible mechanisms for this effect are discussed in section \ref{discussion}. Further clarifications of the robustness of this result with respect to known Subfind issues are found in the Appendix.

The fact that we observe significant differences in ex-situ fraction as a function of parent halo mass suggests a possible explanation for the observation that blue sequence galaxies display lower ex-situ fraction than red sequence galaxies of the same stellar mass (see Figure \ref{repo}); that it is the result of the red sequence galaxies living preferentially in denser regions. To test this hypothesis we undertake the following test; we produce samples of red and blue sequence galaxies in fixed stellar mass bins which have the same halo stellar mass distribution. In practice this means that within each stellar mass bin we randomly sub-select red sequence galaxies (which are more common in the simulations), to match the observed halo mass distribution of the blue galaxies. In the rare cases where no red sequence galaxy of the appropriate stellar mass has the parent halo mass of a particular blue sequence galaxy, both galaxies are removed from their respective samples. Though this significantly reduces the number of objects available in the subset, the halo distributions match one-to-one.

\begin{figure*} 
	\includegraphics[width=0.95\textwidth]{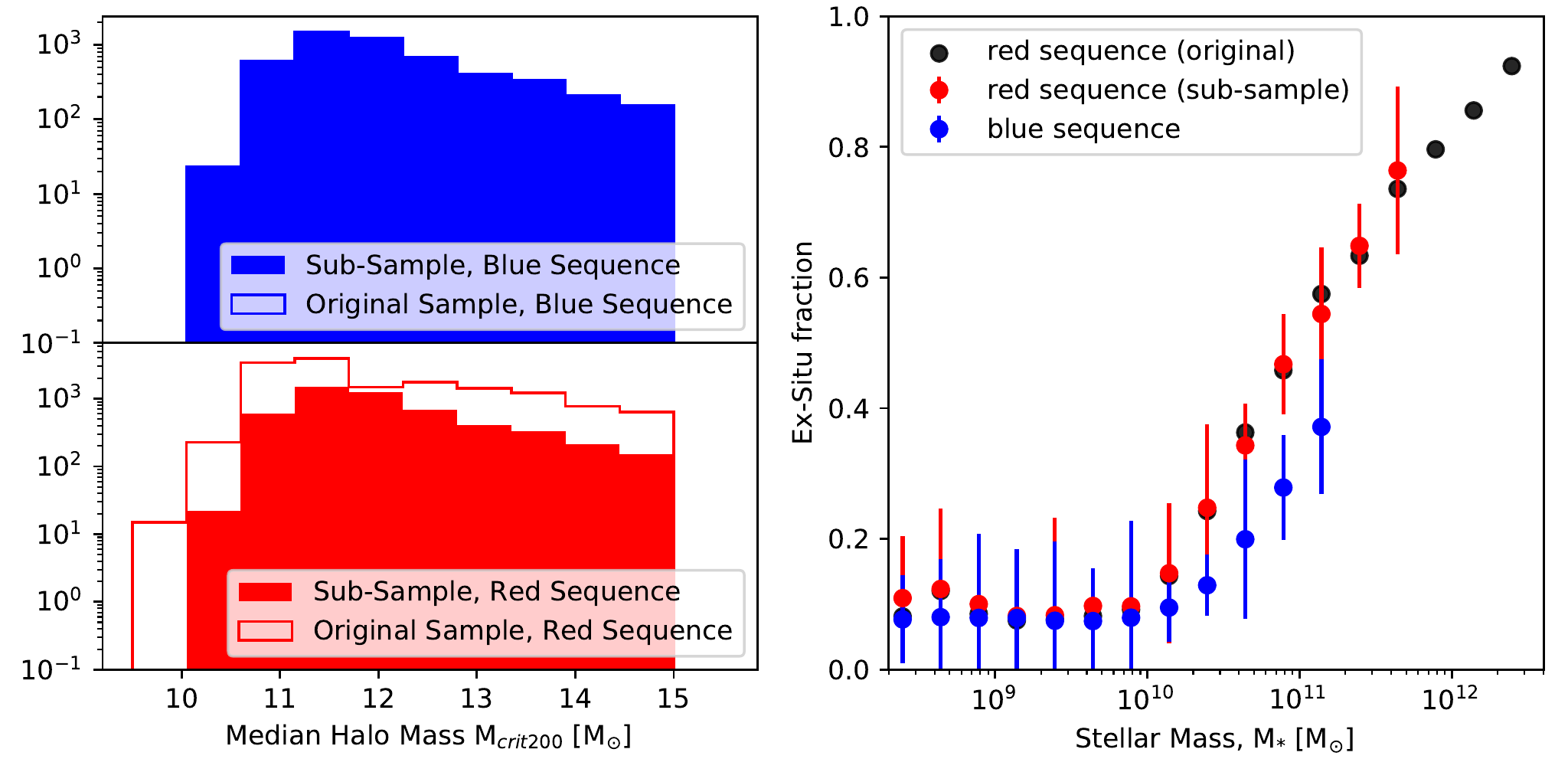}
    \caption{Ex-situ fraction of sub-selected galaxies, such that the halo mass distributions of red and blue sequence samples match. Left panels show original and new sample histograms. There are no residual differences between the two samples. The right-hand panel shows the results of the stellar-mass-ex-situ distribution after the re-sampling with values binned in 0.25dex bins of stellar mass. There is no significant difference between the original data-set of red sequence galaxies, and the re-sampled data-set of red sequence galaxies}
    \label{new_distr}
\end{figure*}

With this subset, we repeat the same analysis as section \ref{ex_sit gal mass} as a function of stellar mass for both galaxy types. This allows us to examine whether samples of early and late type galaxies, matched in mass and halo mass, still display different ex-situ fractions. In Figure \ref{new_distr} we see that the new sub-selected sample shows no preference for higher or lower ex-situ fraction compared to the original sample. Thus persists the systematic trend that early types have a larger ex-situ fraction. We employ 50 Monte-Carlo simulations to estimate error from our random selection, and all original positions are well within these errors. This is possibly a result of the large uncertainties in ex-situ fraction, though the general trends are highly consistent. As a result we can be reasonably sure that the differences in ex-situ fraction between red and blue galaxies at fixed mass do not appear to be driven by systematic differences in halo mass distribution.

\subsection{Ex-situ Fraction across Galaxies’ bodies}
\subsubsection{Ex-situ Fraction as a Function of Radius}
%Figure \ref{fig:amr} shows the Age-Metallicity relation for stars in a high mass galaxy drawn from the Ref-100 simulation. This figure elucidates a number of key facts. First, that for massive galaxies the majority of the stellar mass may have an external origin. Second, that the fraction of ex-situ material increases at larger radii. Third, we see a clear example of inside-out formation in the ages for the in-situ population. We use this division technique to examine ex-situ fractions for other galactic properties.

\begin{figure}
    \includegraphics[width=\linewidth]{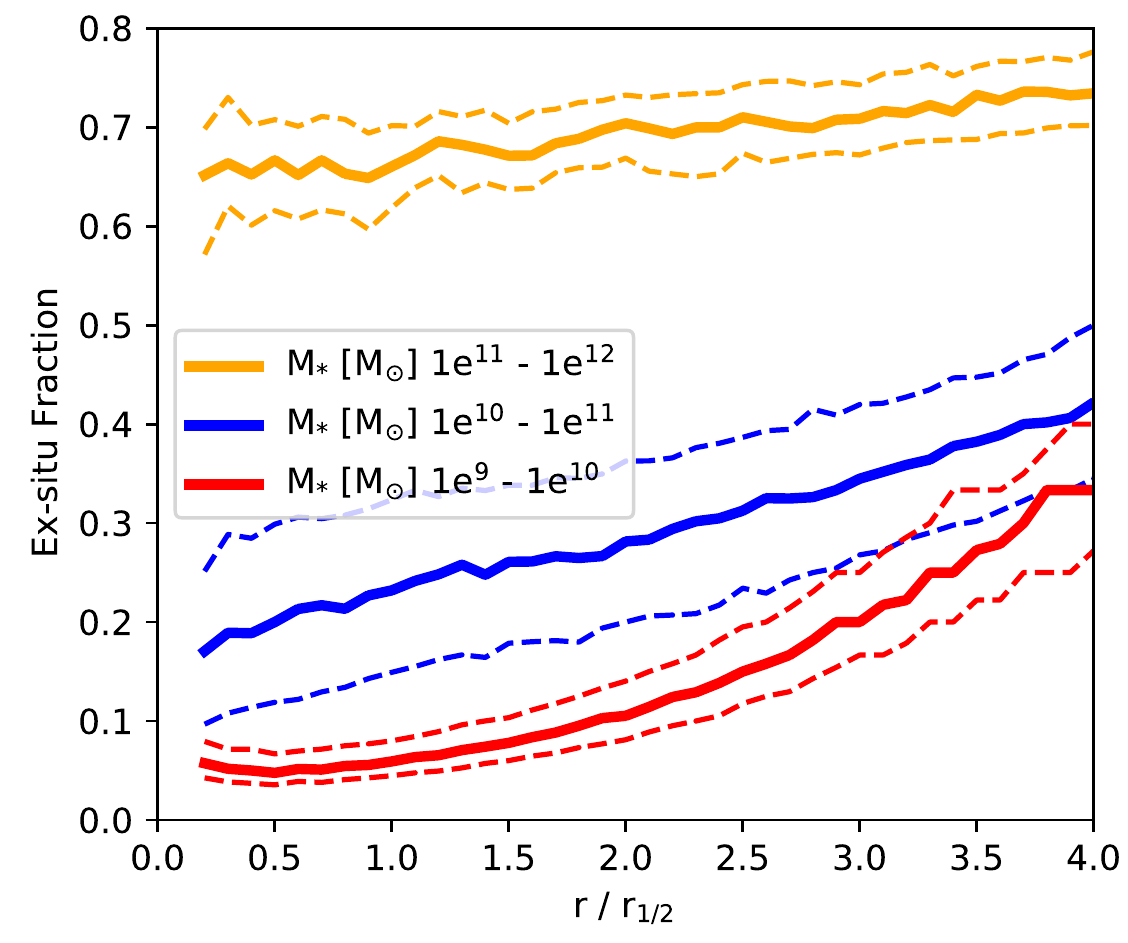}
    \caption{Ex-situ fraction as a function of galactocentric radius, normalised to individual subhalo half mass radius (z=0). Galaxies are binned by total stellar mass. Solid lines show the average ex-situ fraction for a galaxy within the stellar mass bin. Dashed lines show 1$\sigma$ percentile lines of the ex-situ scatter. Mean values are calculated from bins of size 0.1r$_{1/2}$ with 6421 objects in the 10$^{9}$-10$^{10}$M$_{\odot}$ sample, 1418 objects in the 10$^{10}$-10$^{11}$M$_{\odot}$ sample and 263 objects in the 10$^{11}$-10$^{12}$M$_{\odot}$ sample}
    \label{rad_ex_situ}
\end{figure}

%\textcolor{blue}{We were struck by the strength of the location of ex-situ material as a function of radius, especially for massive galaxies of high ex-situ content.} Motivated by this radial division, 
We produce an analysis of all galaxies of > 500 particles as a function of individual subhalo galactocentric distance. When binned by stellar mass, this allows us to see the increase of ex-situ fraction with radius, for each stellar mass bin. This is shown in Figure \ref{rad_ex_situ}. 

From Figure \ref{rad_ex_situ} we see that for less massive galaxies (10$^{9}$ M$_\odot$ - 10$^{10}$ M$_\odot$) not only does the mass fraction of ex-situ stars increase with galactocentric radius, so does the rate of increase. For more massive galaxies (10$^{11}$ M$_\odot$ - 10$^{12}$ M$_\odot$) the ex-situ fraction increase is linear with galactocentric radius, with even the most central regions containing an average ex-situ fraction of > 65\%. This can also be seen in the Illustris simulations, as is shown by \cite{rodriguez2016stellar} in their figure 10.

\subsubsection{Ex-situ Fraction as a Function of Surface Brightness}\label{obsresults}

\begin{figure} 
	\includegraphics[width=\linewidth]{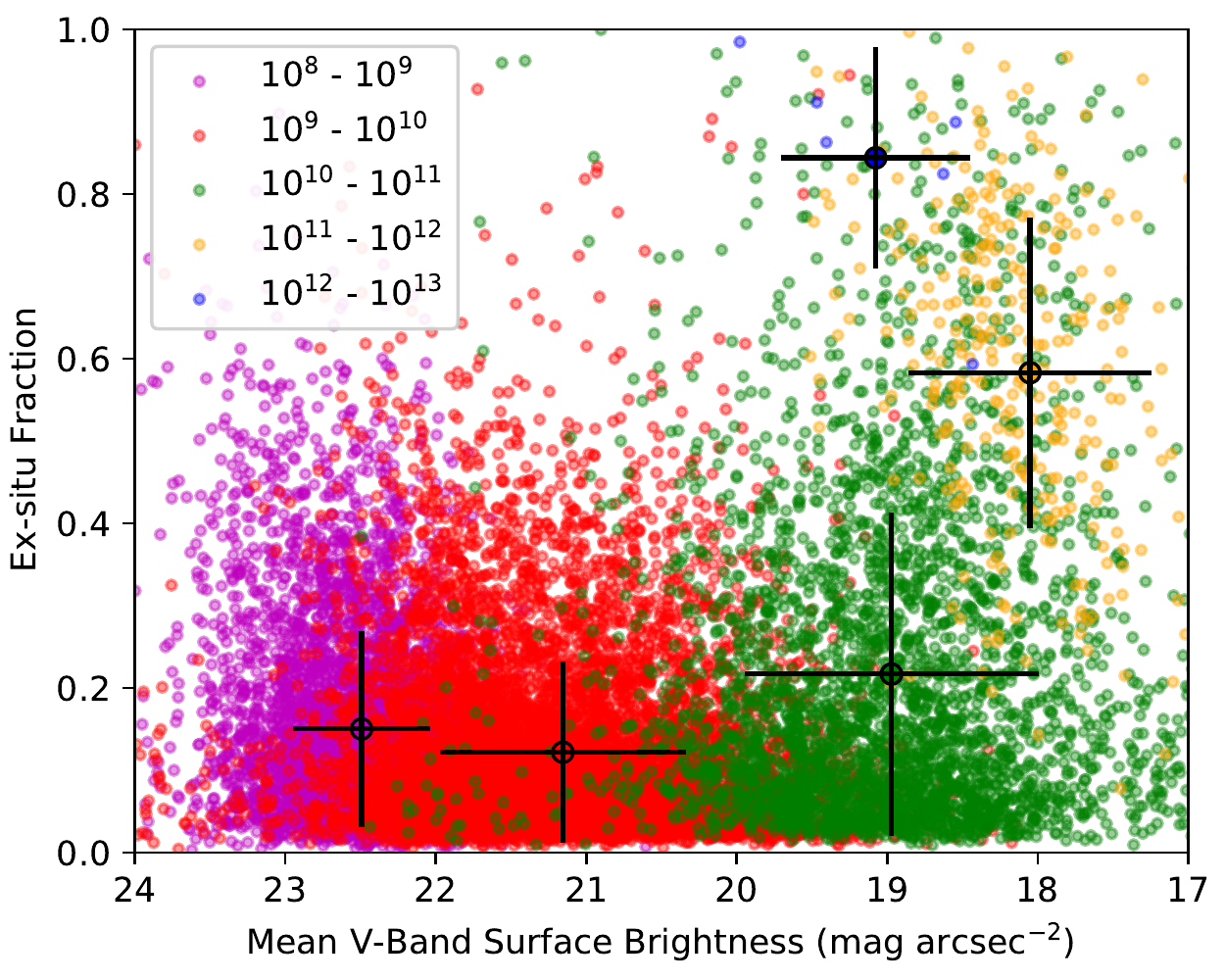}
    \caption{Ex-situ fraction of the z=0 EAGLE galaxies (centrals and satellites) as a function of mean V-band SB within r$_{1/2}$, with separation by mass. The legend shows colour division of stellar mass ranges (M$_{\odot}$). Mean values are overplotted in a larger sized point with 1$\sigma$ standard deviation bars attached. Note that points are plot in order of mass, thus overlapping regions are obscured by higher mass points.}
    \label{situ_mags}
\end{figure}

By linking the ex-situ fraction to the surface brightness (described in section \ref{params}), we create an observationally comparable parameter. This allows us to create predictions of ex-situ fraction which are verifiable observationally. The calculated surface brightness is paired with mass and size parameters of the simulated galaxies, adding a further dimension to the plots. Similarly to the analysis performed to investigate halo mass effects, the galaxies are split by stellar mass. Figure \ref{situ_mags} shows the separation of EAGLE galaxies by mass, along with their average position in the surface brightness/ex-situ fraction plane.

The Figure shows a mostly low ex-situ fraction, whilst surface brightness increases up to a stellar mass of $10^{10} - 10^{11}$ $M_{\odot}$. At this turning point, as mass increases, surface brightness decreases, whilst ex-situ fraction rises. %This is a logical progression, considering our previous observations such as in Figure \ref{smoothed} that those more massive galaxies exhibit greater ex-situ fractions. The larger mass end of the mass size plane is of course dominated by massive ellipticals containing significant old and red stellar populations, and so appear less luminous per unit area compared with smaller bluer galaxies.
This is consistent with our picture of galaxy evolution, in which average surface brightness within r$_{1/2}$ increases with stellar mass initially. Then beyond a stellar mass of $\sim10^{10}-10^{11}M_{\odot}$ ex-situ fraction begins to rapidly increase. As this occurs, the average surface brightness density decreases. This lends weight to the idea that ex-situ stars are preferentially accreted to the outskirts, where they increase the half-light radius, extending the galaxy and thereby reducing the average surface brightness within the effective radius. This can be seen in Figure \ref{rad_ex_situ} where high fractions of ex-situ stars are present at large galactocentric radii compared with the galaxy centres.

We then move to generating predictions for the ex-situ mass fraction as a function of galaxy mass and surface brightness. To ensure the generation of data easily comparable to observable galaxies, our analysis is performed by examining the ex-situ fractions within radial bins containing like surface brightnesses.

Using the method described in Section \ref{params}, magnitudes per square arcsecond (calculated for an assumed distance of 16.5Mpc, the approximate distance to the Virgo cluster) are linked to the ex-situ fractions within the magnitude bins. These values are used to examine the median ex-situ fraction with mass, for constant surface brightnesses.

\begin{figure*} 
	\includegraphics[width=0.95\textwidth]{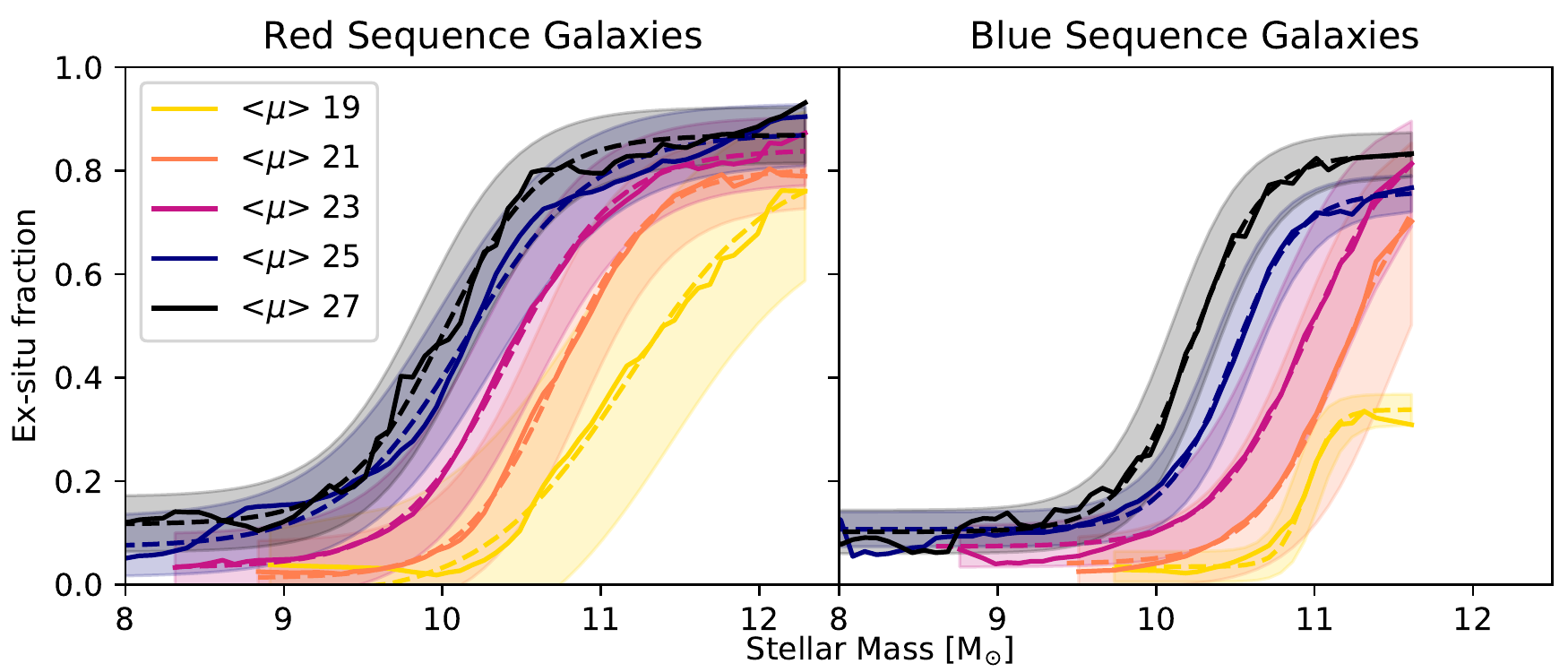}
    \caption{Mass and ex-situ fraction for all z=0 objects, considering only particles within integer magnitudes per square arcsecond projected ellipses in V band. Solid lines show the true data, dashed lines show a fitting function. Sigmoid functions are fit to the data as described in Section \ref{results}. Five different magnitudes-per-square-arcsecond (<$\mu$>) bins are shown here, with additional values tabulated in Table \ref{tab_const}. Shaded regions show the 1$\sigma$ distribution of ex-situ scatter.}
    \label{profile_equal_SB}
\end{figure*}

This is shown in Figure \ref{profile_equal_SB}, where surface brightnesses of magnitudes 19 - 27 per square arcsecond are considered for V band. Sigmoid functions are fit to the data of the form: 
\begin{equation}\label{sb_func}
    \frac{c}{1 + e^{-a\left (x-b \right )}} + d
\end{equation}
where \textit{a,b,c} and \textit{d} are constants tabulated in Table \ref{tab_const}. This allows the expected ex-situ fraction at any stellar mass (8 < log$_{10}$(M$_{\odot}$) < 12.5) to be calculated according to the function at a particular surface brightness. Shaded error bars show the 1$\sigma$ scatter in the ex-situ values within each bin, which are provided as a co-variance matrix of the form shown below, where all elements are tabulated in Table \ref{tab_var}.

\begin{equation}\label{cov_matrix}
    \begin{bmatrix}
     \alpha & \beta &  \gamma & \delta \\ 
    \beta & \epsilon & \zeta & \eta  \\ 
    \gamma & \zeta & \theta & \iota \\ 
    \delta & \eta & \iota  & \kappa
    \end{bmatrix}
\end{equation}

The purpose of this plot, and the tabulated constants is to provide predictions which may be compared directly to forthcoming observational studies. Tabulated constant values are shown in Table \ref{tab_const} for surface brightnesses of 19 - 27 magnitudes per square arcsecond for V band. Values for filters Johnson-Cousins V, B and R; and Sloan g, r and i, are also available as supplementary material online.

\onecolumn
\begin{longtable}{|l|l|l|l|l|}
\caption{Constants for V band surface brightness to ex-situ fraction functions for the stellar mass range 8 \textless log$_{10}$M$_{\odot}$ \textless 12.5 in Equation \ref{sb_func}. Surface brightnesses ($\mu_V$) of 19 - 27 magnitudes per square arcsecond are shown.}
\label{tab_const}\\
\hline
$\mu_V$ & a & b & c & d \\ \hline
\endfirsthead
\endhead
\multicolumn{5}{|l|}{Red Sequence Galaxies} \\ \hline
19 & 2.252 & 11.135 & 0.842 & -0.030 \\ \hline
20 & 3.050 & 10.892 & 0.788 & -0.006 \\ \hline
21 & 3.534 & 10.732 & 0.790 & 0.010  \\ \hline
22 & 3.461 & 10.645 & 0.795 & 0.031  \\ \hline
23 & 3.458 & 10.509 & 0.801 & 0.034  \\ \hline
24 & 3.535 & 10.419 & 0.787 & 0.054  \\ \hline
25 & 3.507 & 10.309 & 0.783 & 0.073  \\ \hline
26 & 3.219 & 10.151 & 0.780 & 0.085  \\ \hline
27 & 2.886 & 10.009 & 0.756 & 0.110  \\ \hline
\multicolumn{5}{|l|}{Blue Sequence Galaxies} \\ \hline
19 & 12.756 & 10.960 & 0.302 & 0.037  \\ \hline
20 & 14.823 & 10.946 & 0.305 & 0.082  \\ \hline
21 & 3.490 & 11.185 & 0.852 & 0.025  \\ \hline
22 & 2.838 & 11.258 & 1.142 & 0.027  \\ \hline
23 & 2.663 & 11.040 & 0.975 & 0.034  \\ \hline
24 & 3.574 & 10.653 & 0.740 & 0.054  \\ \hline
25 & 4.713 & 10.502 & 0.655 & 0.098  \\ \hline
26 & 4.242 & 10.375 & 0.709 & 0.086 \\ \hline
27 & 5.161 & 10.157 & 0.671 & 0.102  \\ \hline
\end{longtable}

\begin{longtable}{|l|l|l|l|l|l|l|l|l|l|l|}
\caption{Constants for V band co-variance matrix (Equation \ref{cov_matrix}) to surface brightness functions for the stellar mass range 8 \textless log$_{10}$M$_{\odot}$ \textless 12.5. Surface brightnesses ($\mu_V$) of 19 - 27 magnitudes per square arcsecond are shown.}
\label{tab_var}\\
\hline
$\mu_V$ & $\alpha$ & $\beta$ & $\gamma$ & $\delta$ & $\epsilon$ & $\zeta$ & $\eta$ & $\theta$ & $\iota$ & $\kappa$ \\ \hline
\endfirsthead
\endhead
\multicolumn{11}{|l|}{Red Sequence Galaxies} \\ \hline
19 & 3.919 & -0.281 & -0.776 & 0.287 & 0.141 & 7.11E-02 & 6.86E-03 & 0.177 & 0.177 & 3.33E-02 \\ \hline
20 & 4.972 & -0.041 & -0.448 & 0.207 & 0.060 & 5.81E-03 & 1.33E-02 & 0.057 & 0.057 & 1.85E-02 \\ \hline
21 & 5.670 & -0.041 & -0.315 & 0.138 & 0.043 & 4.60E-03 & 8.11E-03 & 0.032 & 0.032 & 1.07E-02 \\ \hline
22 & 5.076 & -0.048 & -0.270 & 0.112 & 0.043 & 5.40E-03 & 6.98E-03 & 0.028 & 0.028 & 8.89E-03 \\ \hline
23 & 4.735 & -0.046 & -0.226 & 0.091 & 0.041 & 5.04E-03 & 6.06E-03 & 0.023 & 0.023 & 7.41E-03 \\ \hline
24 & 4.607 & -0.057 & -0.180 & 0.063 & 0.038 & 5.77E-03 & 4.22E-03 & 0.018 & 0.018 & 5.00E-03 \\ \hline
25 & 4.534 & -0.046 & -0.171 & 0.065 & 0.039 & 4.61E-03 & 4.56E-03 & 0.017 & 0.017 & 5.20E-03 \\ \hline
26 & 3.665 & -0.029 & -0.163 & 0.069 & 0.043 & 3.18E-03 & 5.78E-03 & 0.017 & 0.017 & 6.06E-03 \\ \hline
27 & 2.945 & -0.012 & -0.161 & 0.076 & 0.052 & 1.49E-03 & 7.71E-03 & 0.018 & 0.018 & 7.16E-03 \\ \hline
\multicolumn{11}{|l|}{Blue Sequence Galaxies} \\ \hline
19 & 220.933 & -0.256 & -0.769 & 0.219 & 0.011 & 2.72E-03 & 1.19E-03 & 0.008 & -0.002 & 1.88E-03 \\ \hline
20 & 439.691 & -0.331 & -1.018 & 0.259 & 0.012 & 2.97E-03 & 1.15E-03 & 0.009 & -0.002 & 2.01E-03 \\ \hline
21 & 9.661 & -1.224 & -1.852 & 0.169 & 0.225 & 3.03E-01 & -1.35E-02 & 0.449 & -0.030 & 6.45E-03 \\ \hline
22 & 5.746 & -1.500 & -2.413 & 0.129 & 0.497 & 7.72E-01 & -2.43E-02 & 1.245 & -0.048 & 6.31E-03 \\ \hline
23 & 4.264 & -0.985 & -1.273 & 0.100 & 0.321 & 3.75E-01 & -1.46E-02 & 0.475 & -0.028 & 5.56E-03 \\ \hline
24 & 5.745 & -0.343 & -0.457 & 0.065 & 0.061 & 4.38E-02 & 3.19E-04 & 0.055 & -0.007 & 3.32E-03 \\ \hline
25 & 10.459 & -0.202 & -0.326 & 0.057 & 0.030 & 1.37E-02 & 1.45E-03 & 0.021 & -0.003 & 2.48E-03 \\ \hline
26 & 7.935 & -0.172 & -0.303 & 0.063 & 0.034 & 1.39E-02 & 2.02E-03 & 0.024 & -0.005 & 3.34E-03 \\ \hline
27 & 13.924 & -0.125 & -0.306 & 0.080 & 0.026 & 7.37E-03 & 2.39E-03 & 0.018 & -0.004 & 3.63E-03 \\ \hline
\end{longtable}
\twocolumn

\section{Discussion}\label{discussion}
In this study we explore the relation of ex-situ material to galaxy properties, both global and local. In Figure \ref{repo} we plot stellar mass against ex-situ fraction. An increase in ex-situ fraction with stellar mass matches existing expectations (drawn from complementary simulations) about galaxies of larger mass possessing greater fractions of stars accreted from other galaxies. Likewise, less massive galaxies have been shown to have few or even no ex-situ stars \citep{fitts2018no}. When all galaxies are considered, at the highest mass bin limits, the average ex-situ fraction reaches as high as $\approx$84.2\%. This almost identical to the fraction recovered at the highest mass bin by \cite{rodriguez2016stellar} for the Illustris simulations as discussed in section \ref{ex_sit gal mass}.

When comparing to other analyses of the EAGLE simulations, we see small differences between those results and our own. A likely reason for the small differences between analyses of the same simulation are due to differences in ex-situ and in-situ classification. In our method we trace all particles through their entire merger history from formation to $z=0$, resulting in identification of origin. It is possible for a few particles to be mistakenly labelled as ex-situ when swapping between subhaloes during mergers, however this possible over-estimation of ex-situ fraction is negligible and does not impact on the statistics of our sample. 

A second parameter with trends in ex-situ fraction was found to be the total mass of the halo which hosts a satellite galaxy. By forming a sub-sample of satellite galaxies split by the mass of their parent halo we were able to compare the effects of halo mass on ex-situ fraction. As seen in Figure \ref{combo} there is a variation with halo mass such that satellite galaxies within the most massive haloes possess consistently lower ex-situ fractions on average compared to those of the same stellar mass in the next lowest halo mass bin. 
Objects within intermediate mass haloes (M*$_{halo}$>1$\times$10$^{13}$M$_{\odot}$) seem to experience especially efficient accretion of stars onto galaxies.

As a general rule, at fixed stellar mass, satellite galaxies residing in higher mass haloes have lower ex-situ fraction within 100 kpc. This is unexpected when we consider the vastly larger potential wells of larger and likely older haloes. The effect is also visible in similar environmental analyses of other simulations, such as in figure 12 of \cite{pillepich2017first}, in which the IllustrisTNG simulation shows the same effect. Though present in the plot, this is not commented upon, possibly due to the subtle nature of the effect.

Initially the proposed explanation for this effect was as the result of overcooling \cite{crain2015eagle}, in which the most-massive haloes cease to display efficient AGN feedback, leading to artificially high in-situ fractions. This was eliminated as a potential mechanism by running the same analyses, though ignoring stars born after $z=1$. The result shows the same effect as in Figure \ref{combo} in which satellite galaxies within the more massive haloes contain an ex-situ fraction that is lower than galaxies of the same mass in less massive haloes. As such, we can expect there to be more mechanisms responsible than purely overcooling.

A number of mechanisms could be responsible for the effect that satellite galaxies within higher mass haloes have lower ex-situ fraction. One is that because passing velocities in clusters are so high, true mergers are rare, and it becomes more likely that material gets pulled out and added to the intra-cluster light than actually incorporated into another galaxy \citep{moore1996galaxy,makino1997merger}. In contrast, within groups the passing velocities are lower and mergers are more likely \citep{bahcall}. Although we might expect clusters to be built from groups and thus have a similar fraction, it is possible we are instead seeing an influence of survivor bias. This could occur where groups that fell into clusters did so, on average, long ago and ceased merging, whereas surviving groups today have had much longer to continue to merge. 

Evidence for this mechanism can be found in literature \citep[e.g.][]{gu2018spectroscopic} where stellar population models are fit to spectra of intra-cluster light (ICL). The velocity dispersion of the stars indicate that many ICL stars are dissociated from individual galaxies and are instead influenced by the gravitational potential of the cluster as a whole. The authors of the study suggest tidal stripping as a potential formation mechanism for the ICL, where flybys of massive galaxies can expel significant amounts of matter into the ICM.

Another mechanism for this effect could be partly due to differences in galaxy formation efficiency driven by assembly bias. At fixed halo mass, galaxies that form earlier do so more efficiently, because the central density is higher and outflows find it harder to escape \citep{gunn1972infall}. As such, earlier forming galaxies end up with a higher M$_*$/Msub ratio, where Msub is the stellar mass contained in solely subhalos. The satellites of the most massive clusters will have formed the earliest on average, since they are part of higher sigma peaks, and so these could be systematically more efficient at star formation. At fixed M$_*$, we are then looking at systematically lower subhalo mass fractions (as a fraction of the halo mass) as we transition from groups to clusters, and as such you could expect a lower ex-situ fraction.

%We test this second explanation as described in Section \ref{results}, but find that the fixed mass trend still applies across halo masses with respect to VMax. Thus we can safely confirm that there must be a dependence on fixed halo mass, rather than fixed M$_*$.

By separating galaxies based on their half mass radius, we investigate and compare spatial differences regardless of galaxy size. The increase in ex-situ fraction (as seen in Figure \ref{smoothed}) is present both in the central (<r$_{1/2}$) region and outer regions (>2r$_{1/2}$). For both central and outer regions this is the case at all masses, though ex-situ fraction increases negligibly with extent for central regions below a mass of M$_*$ < 5$\times$10$^{8}$M$_{\odot}$. This situation is perhaps to be expected, as higher mass haloes are thought to be assembled through a higher fraction of mergers \citep{maller2006galaxy}.

When looking solely at the central region of galaxies, more diffuse galaxies have accreted a higher fraction of stars than their denser counterparts. This difference is as equally pronounced for the outer regions of the same galaxies, with fractions between 5-30\% greater at all masses where M$_*$ > 5$\times$10$^{8}$M$_{\odot}$. The increase in ex-situ fraction at all masses for the outer regions is indicative of stripped ex-situ stars preferentially remaining in the disk and halo and infrequently migrating to the core. This is supportive of the `two phase' scenario wherein an initial core of in-situ stars accretes stars via mergers and tidal stripping.

Once above M$_*$ > 2$\times$10$^{9}$M$_{\odot}$ the ex-situ fraction of the outer regions quickly surpasses 50\% for the most diffuse galaxies, and all galaxies of mass M$_*$>2$\times$10$^{11}$M$_{\odot}$ contain more than 50\% constituent ex-situ stars, regardless of radius.

Future observational techniques will likely independently estimate ex-situ fractions, with the novel techniques as described in Section \ref{intro}. Figure \ref{profile_equal_SB} converts the simulated data into predictions which may be more readily compared with these future observational estimates.

\section{Conclusions}
We have analysed two EAGLE simulations of volumes 25cMpc$^3$ and 100cMpc$^3$ within the sample stellar mass range of (2$\times$10$^{7}$ - 1.9$\times$10$^{12}$ M$_{\odot}$) for both central and satellite galaxies. From these we have extracted ex-situ fraction information at $z=0$ with respect to spatially resolved galaxy parameters. We have investigated how ex-situ fraction changes with stellar mass, and at fixed mass with changes in half mass radius. Furthermore we have examined ex-situ changes with group/cluster halo mass by separating satellite galaxies by the mass of the group/cluster halo they reside in. Lastly we have used mass to light ratios to determine the expected ex-situ fractions for galaxies at a specific mass and surface brightness value.

Our main findings are summarised as follows:

\begin{itemize}
\item We find that more massive galaxies gain proportionally more stellar mass from ex-situ sources. This is in common with previous findings in literature \citep[e.g. in ][]{oser2010two,qu2016chronicle,rodriguez2016stellar,pillepich2017first}. For the most massive galaxies included in the sample ($>$ 1$\times$10$^{12}$ M$_{\odot}$), ex-situ fraction was found to be, on average, >80\%; and for individual cases at these masses the ex-situ fraction could be as high as $>$90\%.

\item At fixed galaxy mass, when separated by its parent group/cluster halo mass, there is a consistently lower ex-situ fraction for satellite galaxies within the highest mass group/cluster halos. One interpretation of this is that the high passing velocities present in massive clusters disfavours true mergers and stellar accretion. Instead of material being added to a galaxy during a close pass/merger, the material is instead removed from the original galaxy and spread amongst the ICM. Another interpretation is that the effect is the result of differences in feedback efficiency with time, where tightly bound group/cluster halos possess denser circumgalactic medium as a result of less efficient feedback. This in turn causes a higher rate of in-situ formation. %This remains true when considering identical stellar and group/cluster halo mass separations but at fixed galaxy VMax
Similar effects were seen in analysis of a purely central sample, however statistics were too poor to explore this result with any certainty. This remains a point of investigation for larger samples in the future.

\item At all galaxy stellar masses there is an increase in ex-situ fraction with increasing galaxy stellar size, thus showing how at fixed mass, more diffuse galaxies are more likely to contain a greater ex-situ fraction. This supports the idea that physically more extended galaxies are more extended because they have accreted more material preferentially onto their outskirts. This is especially clear in the right hand panels of Figure \ref{smoothed} where bins of the most diffuse, most compact and average galaxies remain separated from the lowest until the very highest mass extent of the simulation.

\item We have also produced predictions for the accreted fraction as a function of galaxy mass and surface brightness. These predictions show estimated ex-situ fractions for the same sample of galaxies, but limited to elliptical isophotes of specific surface brightness values. Combined with stellar mass, these can be used to estimate the ex-situ fraction for observed galaxies of similar surface brightness and mass. This can also be readily compared to observational data processed by new analytic techniques such as recently advanced full spectral fitting methods, for both spatially resolved, and non-spatially resolved galaxies. See Table \ref{tab_const} for the tabulated data for V band. Other bands are also available.
\end{itemize}
\section{Acknowledgements}
The authors gratefully acknowledge insightful and constructive comments from the anonymous referee. This work was completed with support from the ESO Studentship Programme, the Isaac Newton Group of Telescopes Studentship as well as the Moses Holden Studentship, with particular thanks to Patrick Holden. JJD and RAC acknowledge financial support from the
Royal Society.

%%%%%%%%%%%%%%%%%%%%%%%%%%%%%%%%%%%%%%%%%%%%%%%%%%

%All papers should start with an Introduction section, which sets the work in context, cites relevant earlier studies in the field by \citet{Others2013}, and describes the problem the authors aim to solve \citep[e.g.][]{Author2012}.

%Simple mathematics can be inserted into the flow of the text e.g. $2\times3=6$ or $v=220$\,km\,s$^{-1}$, but more complicated expressions should be entered as a numbered equation:

%\begin{equation}
%    x=\frac{-b\pm\sqrt{b^2-4ac}}{2a}.
%	\label{eq:quadratic}
%\end{equation}

%Figures are referred to as e.g. Fig.~\ref{fig:example_figure}, and tables as e.g. Table~\ref{tab:example_table}.

%\begin{table}
%	\centering
%	\caption{}
%	\label{tab:example_table}
%	\begin{tabular}{lccr} % four columns, alignment for each
%		\hline
%		A & B & C & D\\
%		\hline
%		1 & 2 & 3 & 4\\
%		2 & 4 & 6 & 8\\
%		3 & 5 & 7 & 9\\
%		\hline
%	\end{tabular}
%\end{table}

%%%%%%%%%%%%%%%%%%%%%%%%%%%%%%%%%%%%%%%%%%%%%%%%%%



\bibliographystyle{mnras}
\bibliography{biblio} % if your bibtex file is called example.bib

%\newpage
%%%%%%%%%%%%%%%%%%%%%%%%%%%%%%%%%%%%%%%%%%%%%%%%%%

%%%%%%%%%%%%%%%%% APPENDICES %%%%%%%%%%%%%%%%%%%%%
%\appendix

\section{Appendix}
Many substructure identification algorithms, including the Subfind algorithm used here, struggle to identify self-bound substructures against high-density backgrounds. As shown by \cite{knebe2011haloes}, this can result in the spurious "loss" of mass from the outskirts of substructures as their orbit takes them close to the centre of their host halo. Since ex-situ stellar mass is preferentially located in the outskirts of galaxies, this effect has the potential to induce a host halo mass dependence on the \textit{apparent} ex-situ fraction of satellites at fixed stellar mass.

To confirm that this was not a driving effect of the result in Figure \ref{combo}, we restricted a sample to all satellites embedded within halos of mass M$_*$ $>$ 10$^{14}$M$_{\odot}$ (solid red line of Figure \ref{combo}) which would include the most severely impacted galaxies by a potential misidentification issue. By binning galaxies by radial location, as a fraction of the halo radius (r$_{crit200}$) we could investigate if there were discernible differences in ex-situ fraction for satellites closer to the halo centre.

\begin{figure} 
	\includegraphics[width=\columnwidth]{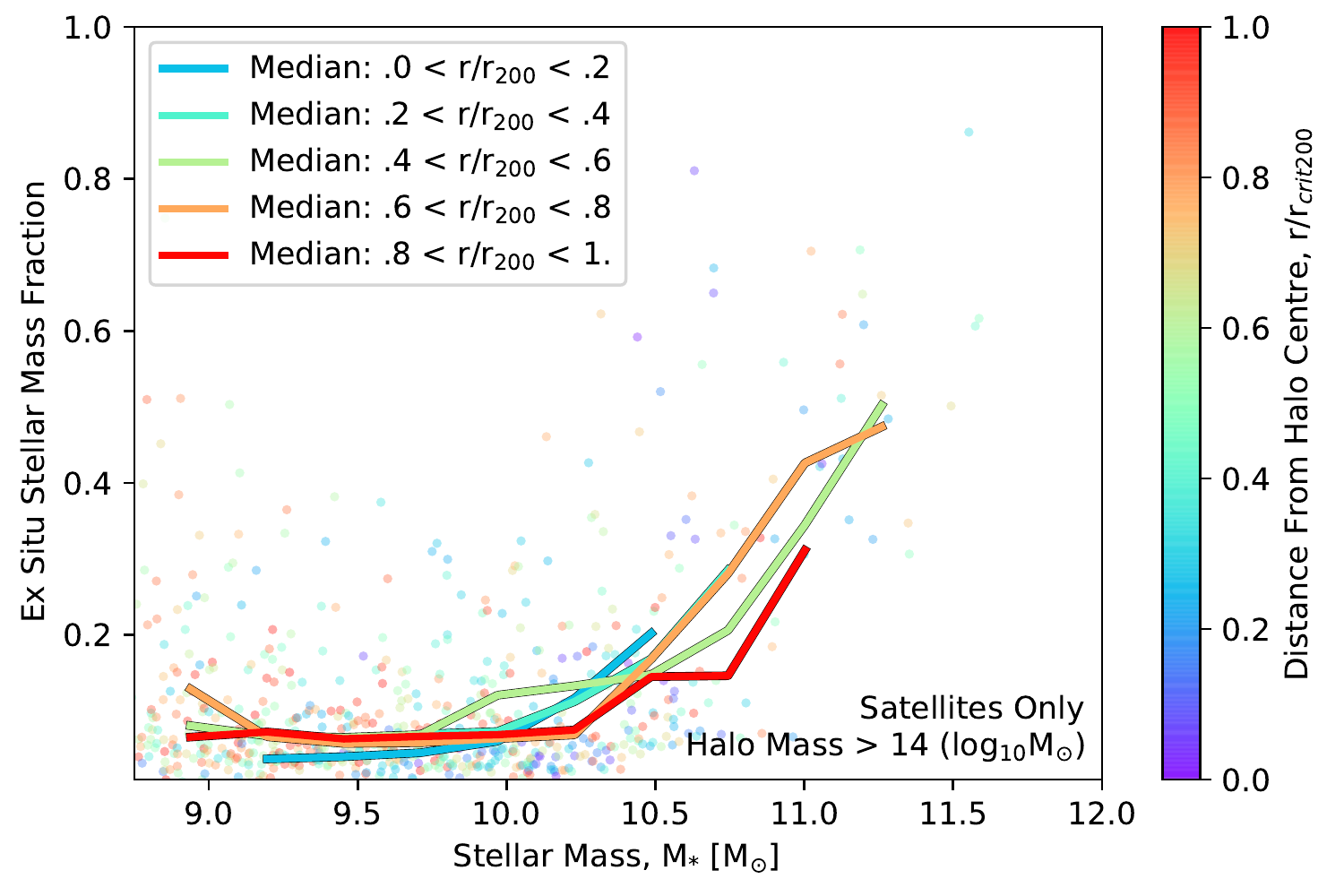}
    \caption{Ex-situ fraction within 100pkpc for z=0 satellite galaxies with $>$500 stellar particles and where M$_*$ $>$ 10$^{14}$M$_{\odot}$. Ex-situ fraction is shown against stellar mass, and split by distance from the halo centre. Individual objects are coloured by the fraction of the distance from the centre of the halo to which they belong, between 0 and 1 r/r$_{crit200}$. Solid lines show the median position of the ex-situ fraction for objects within 5 bins of fractional radial location in the halo, across 0.25 dex wide bins of stellar mass. There are a minimum of 3 objects per bin, and an average of 14 objects per bin.}
    \label{rad_test_01}
\end{figure}

\begin{figure} 
	\includegraphics[width=\columnwidth]{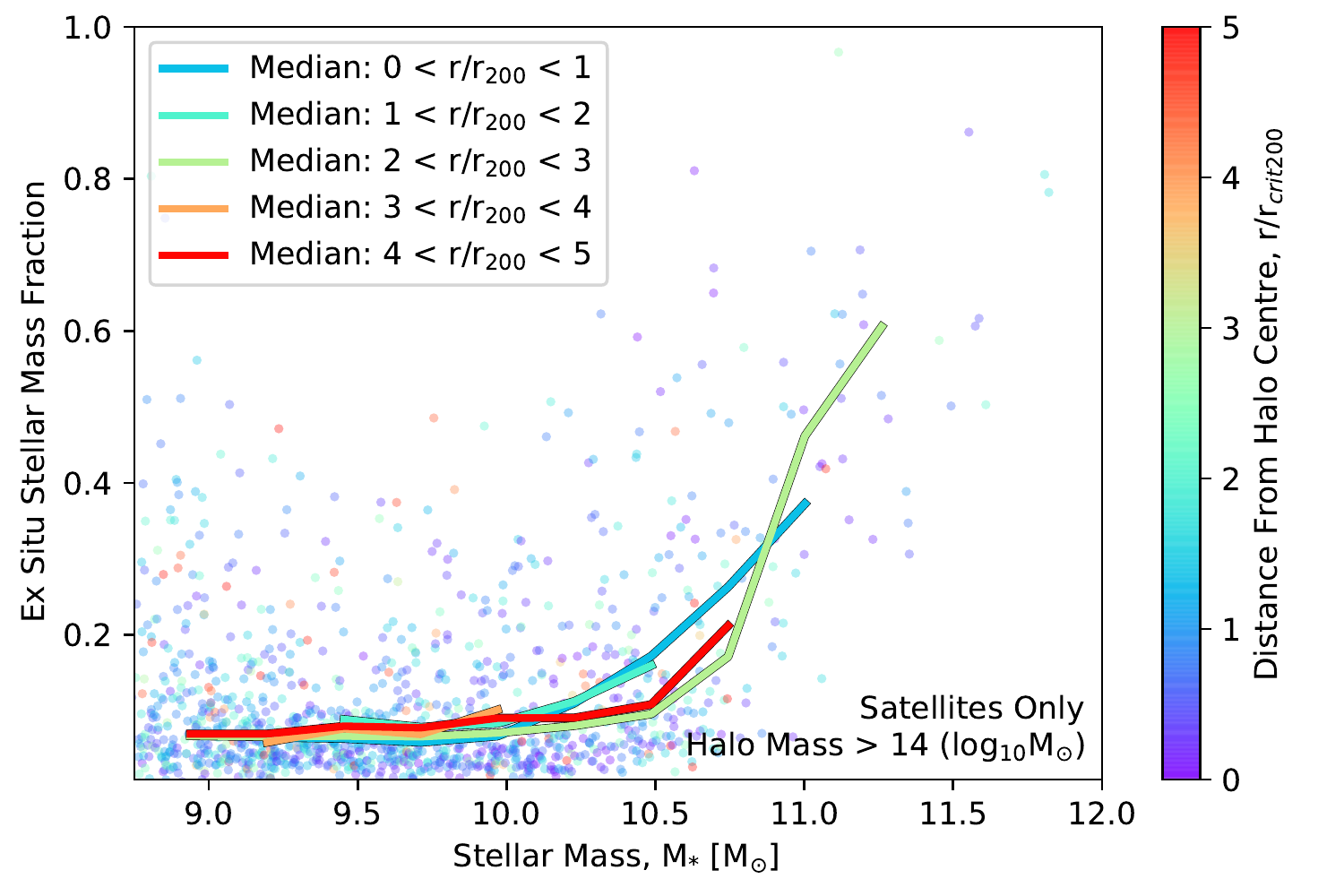}
    \caption{Ex-situ fraction within 100pkpc for z=0 satellite galaxies with $>$500 stellar particles and where M$_*$ $>$ 10$^{14}$M$_{\odot}$. Ex-situ fraction is shown against stellar mass, and split by distance from the halo centre. Individual objects are coloured by the fraction of the distance from the centre of the halo to which they belong between 0 and 5 r/r$_{crit200}$. Solid lines show the median position of the ex-situ fraction for objects within 5 bins of fractional radial location in the halo, across 0.25 dex wide bins of stellar mass. There are a minimum of 3 objects per bin, and an average of 86 objects per bin.}
    \label{rad_test_05}
\end{figure}

As shown in Figures \ref{rad_test_01} and \ref{rad_test_05} there is no impact on the results from any such misidentification effect in the samples used in this study.

This was further confirmed by excluding satellite galaxies in which the extent of the subhalo dark matter component was comparable to that of the stellar component. As a result of the greater extent of the dark matter compared to the stellar component, this Subfind issue would affect the dark matter earlier than for the stars. By excluding the satellites in which (r$_{sm0.5}$ / r$_{dm0.5}$) > 0.55 (where r$_{sm0.5}$ and r$_{dm0.5}$ are the stellar and dark matter half mass radii respectively) we confirmed that we continued to see the effect as seen in Figure \ref{combo}.

\begin{figure} 
	\includegraphics[width=\columnwidth]{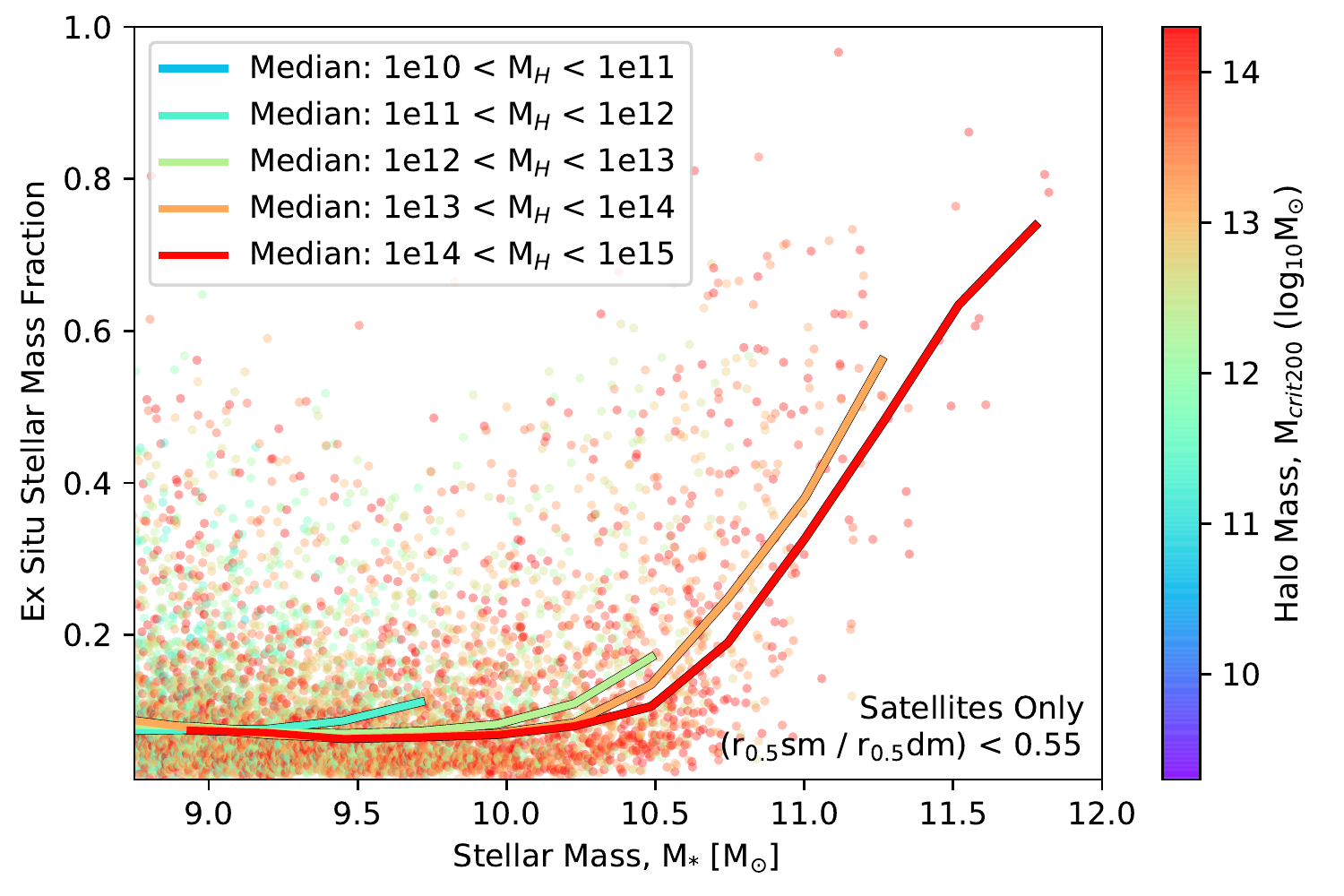}
    \caption{Ex-situ fraction within 100pkpc for all z=0 satellite galaxies with $>$500 stellar particles where r$_{sm0.5}$ / r$_{dm0.5}$) < 0.55. Ex-situ fraction is shown against stellar mass, and split by total halo mass. Individual objects are coloured by the mass of the halo to which they belong. Solid lines show the median position of the ex-situ fraction for objects within 5 bins of parent halo mass, across 0.25 dex wide bins of stellar mass. There are a minimum of 5 objects per bin, and an average of 220 objects per bin.}
    \label{sm_dm}
\end{figure}

This is shown in Figure \ref{sm_dm} where the sample was restricted to satellites in which (r$_{sm0.5}$ / r$_{dm0.5}$) < 0.55 (exclusion of 7.5\% of the original sample). The effect shown in Figure \ref{combo} persists, where at a fixed mass, more massive halos contain satellites of a lower ex-situ fraction. Thus we can confirm that this is most likely a physical result of the simulation, rather than from numerical error.

\label{Append1}

%%%%%%%%%%%%%%%%%%%%%%%%%%%%%%%%%%%%%%%%%%%%%%%%%%

% Don't change these lines
\bsp	% typesetting comment
\label{lastpage}
\end{document}